\documentclass[draftclsnofoot,onecolumn,12pt,twoside]{IEEEtran}
\usepackage[utf8]{inputenc} 
\usepackage[T1]{fontenc}
\usepackage{url}
\usepackage{ifthen}
\usepackage{cite}
\usepackage[cmex10]{amsmath} 

\usepackage{mleftright}       
\mleftright                   

\usepackage{graphicx}         
\usepackage{booktabs}         

\usepackage{bbm}
\usepackage{svg}
\usepackage{upgreek}
\usepackage[linesnumbered]{algorithm2e}
\usepackage[shortlabels]{enumitem}
\usepackage{tikz}
\usepackage{lipsum,adjustbox}
\usetikzlibrary{calc,positioning}

\usepackage{amssymb}
\usepackage{amsfonts} 
\usepackage{tcolorbox}
\usepackage{enumitem}

\newcommand{\rl}{L}
\newcommand{\noofreads}{K} 
\newcommand{\bl}{n}
\newcommand{\cover}{c}
\newcommand{\eras}{\perp}
 \newcommand{\ssechannel}{\mathsf{SSE}(\delta)}
 \newcommand{\bec}{\mathsf{BEC}(\delta)}
\newcommand{\vz}{\underline{z}}
\newcommand{\vx}{\underline{x}}
\newcommand{\vy}{\underline{y}}
\newcommand{\vynoerasures}{\tilde{\underline{y}}}
\newcommand{\caly}{{\cal Y}}
\newcommand{\calynoerasures}{\tilde{\caly}}
\newcommand{\expect}{\mathbb E}
\newcommand{\indicatorRV}{\mathbb I}

\newcommand{\vu}{\underline{u}}
\newcommand{\vv}{\underline{v}}
\newcommand{\vomega}{\underline{\omega}}
\newcommand{\start}[1]{{\mathsf S}({#1})}

\newcommand{\suff}[2]{\texttt{suffix}(#1, #2)}
\newcommand{\pref}[2]{\texttt{ prefix}(#1, #2)}

\newcommand{\size}[1]{{\ell}_{ue}(#1)}
\newcommand{\length}[1]{{\ell}(#1)}

\newcommand{\Var}{\mathrm{Var}}
\newcommand{\Cov}{\mathrm{Cov}}
\newcommand{\normrl}{\bar{\rl}}
\newcommand{\gbar}{\bar{G}}
\newcommand{\tth}{\texttt{th}}
\newcommand{\lambdaue}{\omega^t_1}

\newcommand{\avglambdaue}{\bar{\omega}^t_1}
\newcommand{\valueofavglambdaue}{\mathsf{w}}

\newcommand{\normsteplambdaue}{d}
\newcommand{\intindex}{k}
\newcommand{\erasprob}{\delta}
\newcommand{\overlaplength}{L_1}
\newcommand{\slackforavglambdaue}{p}
\newcommand{\expon}{\mathsf{e}}
\newcommand{\bigo}{\mathsf{O}}

\newtheorem{lemma}{Lemma}

\newtheorem{remark}{Remark}
\newtheorem{claim}{Claim}
\newtheorem{definition}{Definition}
\newtheorem{theorem}{Theorem}

\RestyleAlgo{ruled}

\begin{document}
\title{On Achievable Rates for the Shotgun Sequencing Channel with Erasures} 
\author{%
 \IEEEauthorblockN{Hrishi Narayanan, Prasad Krishnan, Nita Parekh}
 \IEEEauthorblockA{\\IIIT Hyderabad\\
                   Emails: \{hrishi.narayanan@research., prasad.krishnan@, nita@\}iiit.ac.in}
}

\maketitle


\begin{abstract}
   In shotgun sequencing, the input string (typically, a long DNA sequence composed of nucleotide bases) is sequenced as multiple overlapping fragments of much shorter lengths (called \textit{reads}). Modelling the shotgun sequencing pipeline as a communication channel for DNA data storage, the capacity of this channel was identified in a recent work, assuming that the reads themselves are noiseless substrings of the original sequence. Modern shotgun sequencers however also output quality scores for each base read, indicating the confidence in its identification. Bases with low quality scores can be considered to be erased. Motivated by this, we consider the \textit{shotgun sequencing channel with erasures}, where each symbol in any read can be independently erased with some probability $\delta$.  We identify achievable rates for this channel, using a random code construction and a decoder that uses typicality-like arguments to merge the reads. 
\end{abstract}

\section{Introduction}
Due to its exceptional stability, DNA molecules show great promise as a medium for long term storage of digital data. In the past decade, there has been considerable academic interest in studying and developing DNA as an archival storage medium. This includes practical demonstration of DNA as a storage medium \cite{carmean_hybrid_moelcular_electronic_computing} and potential automation of the storage pipeline \cite{takahashi_end_to_end}; characterization of errors in the DNA storage pipeline \cite{heckel_characterization_DNA_storage}, including in silico simulation of errors due to decay of the molecule \cite{grass_robust_chemical_presevation}; developing algorithms for sequencing of DNA molecules and alignment of reads \cite{bresler_optimal_assembly, shomorony_info_optimal_genome_assembly}; developing error-correction codes for DNA storage \cite{nassirpour_reassembly_chop_and_shuffle, bar-lev_adverserial_torn_paper, lenz_coding_over_sets}; and, information theoretic characterization and analysis of DNA storage \cite{motahari_info_theory_dna-shotgun_sequencing, shomorony_limits_adversarial_erasure, ravi_coded_ssc, levick_achieving_capacity_linear_codes, heckel_fundamental_limits, shomorony_capacity_results_noisy_shuffling,  shomorony_torn_paper, ravi_torn_paper_lost_pieces, lenz_upper_bound_capacity_of_DNA_storage, lenz_achieving_capacity_of_DNA_storage,Shomo_Heckel_TIT_ShufflingSampling_2021}. 

These advancements, as well as the academic and industrial interest, in DNA storage have been fuelled by the developments in sequencing techniques, which reduced costs and increased the feasibility of such storage systems. One such significant development is that of the \emph{high-throughput shotgun sequencing} pipeline. Rather than sequencing the long DNA sequence in its entirety, which is very expensive and often impractical, in shotgun sequencing, multiple copies of the DNA molecule are first created, and then broken into fragments using restriction enzymes. The fragments are generally much shorter in length than the original molecule. These fragments are then sequenced by the shotgun sequencer, resulting a collection of \textit{reads}. The reads are then subsequently aligned, by mapping the overlaps between them, to reconstruct the original sequence.

The work by Lander and Waterman \cite{lander_genomic_mapping} was one of the early ones to analytically model the problem of DNA sequence assembly and derive various limits on the parameters that govern the sequencing, such as the read length and the coverage, for reliable reconstruction. 
Building on the model developed in \cite{lander_genomic_mapping}, Motahari \emph{et al.} \cite{motahari_info_theory_dna-shotgun_sequencing} studied the shotgun sequencing channel from an information theoretic perspective. In this work, the length of reads was considered to scale as $\normrl \log{\bl}$, where $\normrl$ is a fixed constant and $\bl$ is the length of the input sequence $\vx$, which was assumed to be generated uniformly at random from all possible quaternary sequences. The work \cite{motahari_info_theory_dna-shotgun_sequencing} demonstrated some necessary and sufficient conditions on $\normrl$, as well as a parameter known as the \textit{coverage depth} $c$ (which captures the average number of times any position in $\vx$ occurs in the collection of reads), for the reconstruction of $\vx$ in the asymptotic regime, i.e., when $\bl \to \infty$. 

The approach taken in \cite{motahari_info_theory_dna-shotgun_sequencing} also proved useful in studying the fundamental limits of DNA data storage, where the goal is to find the \textit{capacity} of the DNA sequencing channel, i.e., the largest normalized size of any collection of input sequences which can be decoded with vanishing error probability when transmitted through the DNA sequencing channel. For instance, the works \cite{heckel_fundamental_limits, Shomo_Heckel_TIT_ShufflingSampling_2021}  identified the capacity of the \emph{Sampling-Shuffling Channel}, where data is stored as a set of short DNA strands of equal length, and reads are obtained by sampling (with replacement) from this set. The capacities of noisy versions of this channel, with errors and erasures, were also presented in \cite{Shomo_Heckel_TIT_ShufflingSampling_2021}. 
The capacity of the Shotgun Sequencing Channel (with binary-valued inputs) was presented in \cite{ravi_coded_ssc}. In this work \cite{ravi_coded_ssc}, $\noofreads$ reads of fixed length $\rl = \normrl \log{\bl}$ are obtained by uniformly sampling their starting positions from the indices of the input string $\vx$. The capacity of this channel was shown to be $(1-e^{-c(1-1/{\normrl})})$, where $c=\frac{\noofreads\rl}{\bl}$ denotes the \emph{coverage depth} (the expected number of times any position of $\vx$ occurs in the collection of reads). The achievability in \cite{ravi_coded_ssc} is obtained essentially via a random coding argument, where the decoding algorithm used typicality-like techniques to reconstruct the input codeword from the reads. A matching converse result is obtained using a genie-aided argument, using the idea of `islands', which refers to maximal collections of reads which overlap with each other. Results from an earlier work \cite{ravi_torn_paper_lost_pieces} on the so-called \textit{torn paper channel} were used in proving the converse in \cite{ravi_coded_ssc}. A related series of works \cite{ElischoGabrysYaakobi_TIT_RepeatFree,MarkovichYaakobi_TIT_ReconstructSubstringSpect} considers the string reconstruction problem from its \textit{$L$-multispectrum}. The \textit{$L$-multispectrum} of a string is the multiset of all possible $L$-length (contiguous) substrings of the string, taken from every possible starting position. The works \cite{ElischoGabrysYaakobi_TIT_RepeatFree,MarkovichYaakobi_TIT_ReconstructSubstringSpect} effectively derive results on the capacity of channels in which the output of the channel is either (a) the complete $L$-multispectrum, or (b) an incomplete $L$-multispectrum, missing some $t$ substrings, or (c) the $L$-multispectrum with some fixed number of $L$-length substrings having substitution errors. For such models, capacity results and explicit codes are obtained for the case when $L=a\log \bl$ for some $a>1$. Specifically, the work \cite{MarkovichYaakobi_TIT_ReconstructSubstringSpect} considers the code design and capacity analysis of the incomplete $L$-multispectrum channel for a worst-case subset and for specific choices of $t$. The output of the channel model in \cite{ravi_coded_ssc} is also a subset of the $L$-multispectrum, however the $K$ reads (substrings of length $L$) are generated by sampling their starting points in a uniformly random fashion, modelling the coverage through the parameter $c$. Thus, the channels of \cite{ravi_coded_ssc} and \cite{ElischoGabrysYaakobi_TIT_RepeatFree,MarkovichYaakobi_TIT_ReconstructSubstringSpect} are related but distinct. 

In many modern sequencers, each nucleotide that is sequenced would be accompanied by a quality score, for instance Phred scores\cite{li_adjust_quality_score_from_alignment}, corresponding to the probability that the base call for that nucleotide is erroneous. 
By applying a suitable threshold on the score value, we can treat base calls with scores below the threshold as erasures. 
Considering the availability of such quality scores, recent works on the information-theoretic characterization of the DNA storage channel (for instance, \cite{shomorony_it_fundamentals_DNA_storage, seiyun_erasure_shuffling}) have considered the low-quality base calls as \textit{erasures} in the reads.

 In the present work, we consider the shotgun sequencing channel \emph{with erasures}, motivated by the need to incorporate the availability of quality scores of the bases sequenced. 
 The model is similar to that in \cite{ravi_coded_ssc}, with the addition being that each symbol in each read is assumed to be erased with probability $\delta$. We denote this channel as $\ssechannel$ (thus, $\mathsf{SSE(0)}$ is the channel considered in \cite{ravi_coded_ssc}). In this work, we obtain an achievability result for the channel $\ssechannel$ using random coding and typicality-like decoding arguments, thus showing a lower bound on its capacity. The mathematical techniques adopted to show the achievability result generally follow those in \cite{ravi_coded_ssc}. However, there are differences that arise. In some parts, we are able to simplify the analysis as compared to \cite{ravi_coded_ssc}. In others, the analysis is more complicated, owing to the fact that we have to account for the erasures in the reads. Section \ref{section:Channel description} provides the formal channel model and the main result of the paper, along with comparisons to existing work. Section \ref{sec:achievability} gives the coding scheme and the proof of the achievability. The paper concludes with some remarks in Section \ref{sec:conclusion}.

 \emph{Notation:} 
 In this work, ordered tuples or strings are denoted with underlines, such as $\vx$. We denote the set of integers $a,a+1,\hdots,b$ as $[a:b]$. The set of integers $[1:b]$ is denoted as $[b]$. For an event $A$, the indicator random variable associated with the event is denoted by $\indicatorRV_A$. The probability of an event $A$ is denoted by $\Pr(A)$. The complement of an event $A$ is denoted by $\overline{A}$. For two events $A,B$, we write $\Pr(A,B)$ for the probability $\Pr(A\cap B)$. For a set $S$, the set of finite length strings with symbols from $S$ is denoted by $S^\star$. All logarithms are in base $2$.

\section{Channel Description and Main Result}
\label{section:Channel description}

\begin{figure*}[tbh]
\centering
\includegraphics[width=\textwidth]{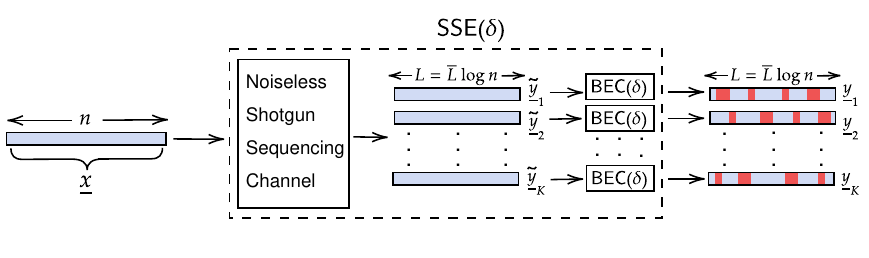}
\caption{The Shotgun Sequencing Channel with Erasures ($\ssechannel$). The collection ${\cal \tilde{Y}} = \{\tilde{\vy}_1, \tilde{\vy}_2, \cdots, \tilde{\vy}_{\noofreads} \}$ may be visualized as the output of the Shotgun Sequencing Channel \cite{ravi_coded_ssc}, and ${\caly} = \{\vy_{1}, \vy_{2}, \cdots, \vy_{\noofreads} \}$ is the output of $\ssechannel$, after bits in each read are erased (indicated in bold/red) with probability $\erasprob$.}
\label{fig:SSEchannel}
\end{figure*}

We follow the description and terminology similar to those in \cite{ravi_coded_ssc}, as the present work essentially extends the achievability result in \cite{ravi_coded_ssc} to the erasure scenario. 

The channel takes an $\bl$-length binary\footnote{The symbols in a DNA sequence take values in a  quaternary set, but for simplicity we assume the symbols to be binary. Our results can be easily extended to the quarternary case.} string $\vx=(x_1,\hdots,x_\bl) \in \{0,1\}^\bl$ as input, corresponding to a message $W \in [2^{\bl R}]$   chosen at random. The output of the channel can be envisioned as a concatenation of two stages, as shown in Fig. \ref{fig:SSEchannel}. Firstly, the channel samples $\noofreads$ substrings of length $\rl$, from $\vx$. We denote these by a multiset $\calynoerasures=\{\vynoerasures_1,\hdots,\vynoerasures_K\}$. Each read $\vynoerasures_i$ is obtained by first selecting a position $\start{\vynoerasures_i}$ uniformly at random from $[\bl]$, and then taking the $\rl$-length (contiguous) substring from the position $\start{\vynoerasures_i}$ onwards, i.e., $\vynoerasures_i=(x_{\start{\vynoerasures_i}},\hdots,x_{\start{\vynoerasures_i}+\rl-1})\in\{0,1\}^\rl$. When $\start{\vynoerasures_i}>\bl-\rl+1$, similar to the circular DNA model in \cite{motahari_info_theory_dna-shotgun_sequencing}, we assume that the substring is obtained in a cyclic wrap-around fashion, for ease of analysis. Thus, $\calynoerasures$ can be thought of as the output of a noise-free shotgun sequencing channel (as in \cite{ravi_coded_ssc}), when the input is $\vx$. In the second stage, each read $\vynoerasures_i$ is assumed to pass through a binary erasure channel with erasure probability $\erasprob$ (denoted by $\bec$), thus erasing each position in $\vynoerasures_i$ with probability $\erasprob$ independently, to obtain $\vy_i\in\{0,1,\eras\}^\rl$, where $\eras$ denotes an erasure. The multiset of these reads, denoted as $\caly=\{\vy_i:i\in[\noofreads]\}$, is the output of the channel. Note that the start positions are unaltered, i.e., $\start{\vy_i}=\start{\vynoerasures_i}, \forall i$. We denote this shotgun sequencing channel with erasures as $\ssechannel$. A rate $R$ is said to be \textit{achievable} on $\ssechannel$ if the message $W$ can be reconstructed from $\caly$ using some decoding algorithm with a probability of error that is vanishing as $\bl$ grows large. The capacity of $\ssechannel$ is then defined as $C_{\ssechannel}\triangleq\sup\{R:R~\text{is achievable}\}$.

The expected number of times a coordinate of $\vx$ (say the $j^\tth$ coordinate) is sequenced in the first stage is called the coverage depth, denoted as $c$. Thus, $c\triangleq \expect(\sum_{i=1}^\noofreads\indicatorRV_{\{j\in [\start{\vy_i}:\start{\vy_i}+L-1]\}}).$  A simple calculation reveals that 
\begin{align}
\label{eqn:valueofc}
    c = \frac{\noofreads\rl}{\bl}
\end{align}
We assume that the length of each read is 
$
\rl=\Theta(\log{\bl})\triangleq\normrl\log{\bl}. 
$
for some positive $\normrl$. As in \cite{ravi_coded_ssc}, we study the regime where $c$ and $\normrl$ are some positive constants. Thus, in our regime,
$
\noofreads=\frac{c\bl}{\normrl\log{\bl}}=\Theta\left(\frac{\bl}{\log{\bl}}\right).     
$ 

 In this work, we obtain a lower bound for $C_{\ssechannel}$, by demonstrating an achievability scheme via a random code construction and a decoder which uses typicality-like techniques for estimating the true message. The main result in this work is the following.

\begin{theorem}
\label{thm:main}
    Let $c$ and $\normrl$ be the parameters of $\ssechannel$ such that $c>0$ and $\normrl(1-\erasprob)>1$. Let $\alpha=c/(\normrl(1-\erasprob))$. The rate $R$ is achievable on $\ssechannel$ if 
    \begin{align}
    \label{eqn:achievablerates}
     R<\left(1- e^{-c(1-\erasprob)}\right) - (1-\delta)\left(e^{-c\left(1-\frac{1}{\normrl(1-\delta)} \right)} - e^{-c}\right).
    \end{align}
\end{theorem}
\begin{remark}
Note that at $\erasprob = 0$, the R.H.S. of (\ref{eqn:achievablerates}) is $(1-e^{-c(1-1/{\normrl})})$, which is the capacity of the shotgun sequencing channel (without erasures) as shown in \cite{ravi_coded_ssc}. 
\end{remark}

\begin{remark}
    In the proof of Theorem \ref{thm:main}, we show that, for every $d > 0$, rate $R$ is achievable on $\ssechannel$ if 
    \begin{align}
    \label{eqn:beta(d)theoremeqn}
        R<\left(1- e^{-c(1-\erasprob)}\right) - \frac{ce^{-c}}{\normrl} - \upbeta(d),
    \end{align}
    where $\upbeta(d)= \frac{\normsteplambdaue}{(1-\erasprob)} \cdot\left(\frac{c}{\normrl}\right)^{2} \cdot e^{-c}\cdot \left(\frac{e^{\alpha\normsteplambdaue}(e^{\alpha}-1)}{(e^{\alpha\normsteplambdaue}-1)} - \frac{e^{2\alpha \normsteplambdaue}\left(\left(e^{\alpha(1+\normsteplambdaue)} - e^{\alpha}\right) - \normsteplambdaue\left(e^{\alpha(1+\normsteplambdaue)} - 1\right)\right)}{\left(e^{\alpha\normsteplambdaue} - 1\right)^2}\right).$ As $d\to 0$, the R.H.S. of (\ref{eqn:beta(d)theoremeqn}) reduces to the R.H.S. of (\ref{eqn:achievablerates}) in Theorem \ref{thm:main}. Simulation results show that the value of $\upbeta(d)$ decreases as $d$ decreases, thereby hinting that the R.H.S. of (\ref{eqn:achievablerates}) is the largest possible achievable rate that can be obtained via proving (\ref{eqn:beta(d)theoremeqn}). However, we are currently unable to prove this analytically. 
\end{remark}


\begin{figure}[htbp]
\centering
\includegraphics[width=\columnwidth]{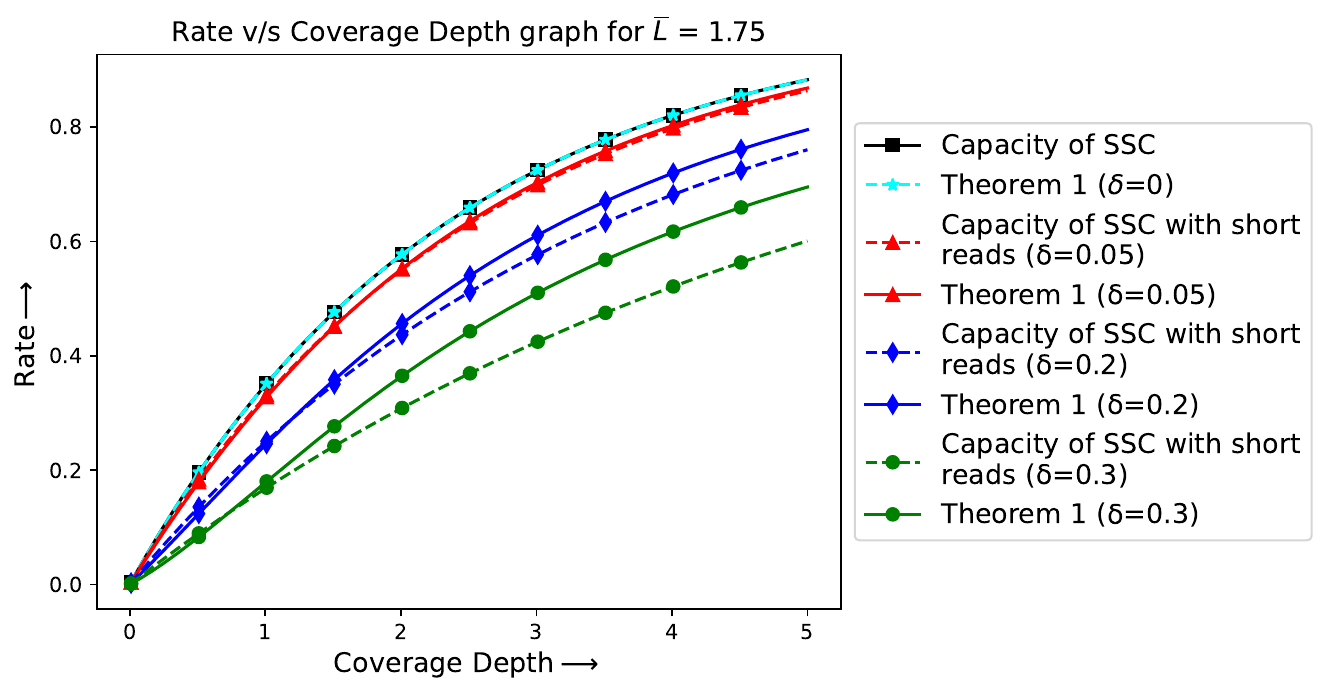}
\caption{The plot shows the rates from Theorem \ref{thm:main}, with $\normrl=1.75$, as the coverage depth $c$ varies, for $\delta=0, 0.05, 0.2, $ and $0.3$. We compare these with the capacity of the shotgun sequencing channel (denoted by SSC) from \cite{ravi_coded_ssc} with read lengths $\normrl(1-\delta)\log\bl$.}
\label{fig:plot}
\end{figure}

Fig. \ref{fig:plot} plots the upper bound for the achievable rate for $\ssechannel$ from Theorem \ref{thm:main}, for $\erasprob\in\{0,0.05,0.2,0.3\}$, against varying values for the coverage depth $c$. The parameter $\normrl$ is fixed as $1.75$ (thus satisfying the requirement $\normrl(1-\erasprob)>1$, for all chosen $\erasprob$).
As a note of comparison, we plot the SSC capacity from \cite{ravi_coded_ssc}, with shortened reads of size $\normrl(1-\erasprob)\log\bl$ (note that the read length is $\normrl\log\bl$ in $\ssechannel$). We observe that this short-read SSC capacity is larger than our bound from Theorem \ref{thm:main}, when $c$ is small (roughly, $c<1$), whereas it is progressively smaller compared to our bound, as $c$ increases (for given $L$, this means $K$ increases). We will now remark on why this may be the case. Firstly, we observe that, due to the length being shorter, the number of reads $K$ for the SSC is larger than $K$ for the $\ssechannel$, for any specific $c$. Despite this, for larger values of $c$, some information about the relative positions of the bits in the transmitted sequence is likely lost by the SSC, as the read length is shorter. As a result, in the case of $\ssechannel$, each contiguous string obtained after merging the reads as per the overlaps tends to be longer and the number of such strings will be smaller, in comparison to SSC with shorter reads. Hence,  the $\ssechannel$ channel is probably able to preserve the information about the relative positions better, due to the longer reads, in spite of the erasures and lesser $K$. In the small $c$ regime, there are likely too few reads in $\ssechannel$ to see this advantage. Instead, due to $K$ being less, the reconstructed sequence in $\ssechannel$ likely has many unrecoverable bits, compared to the SSC channel (in spite of its shortened reads). This arguably leads to the behaviour seen in Fig. \ref{fig:plot}. 

\section{Achievability (Proof of Theorem \ref{thm:main})}
\label{sec:achievability}

We use a random coding argument to show the achievability of the rate as in Theorem \ref{thm:main}. We outline the main components of our code design below. While these share similarities to the techniques in \cite{ravi_coded_ssc}, the decoding algorithm and the proof arguments are more complex, owing to consideration of the reads with erasures. 
\subsection{Outline of the Coding Scheme}
\label{subsec:outline}
\begin{itemize}[leftmargin=*]
    \item \underline{Codebook:} A codebook with $2^{\bl R}$ codewords, denoted as ${\mathcal{C}}=\{\vx_1, \vx_2, \cdots, \vx_{2^{\bl R}}\}$, is generated by picking each symbol of $\vx_j$ independently and uniformly at random from $\{0,1\}$, for each $j\in[2^{nR}]$. 
    \item \underline{Encoder:} To communicate the message $W$ (chosen uniformly at random from $[2^{nR}]$) through the channel $\ssechannel$, the encoder communicates the codeword $\vx_{W} \in {\mathcal{C}}$ through the channel. 
    The output $\caly$, a set of reads as described in Section \ref{section:Channel description}, is generated post-sequencing. 
    \item \underline{Decoder:} The decoding algorithm we propose takes as input the collection of reads $\caly$ and generates an estimate $\hat{W}$ of the transmitted message, or a failure. We briefly describe the process of obtaining the estimate $\hat{W}$ from $\caly$. The decoder proceeds in three phases. In the first phase, which we call the \textbf{merge phase}, the decoder first implements a merging process of the reads. Such a merging process will be run for all possible orderings of the reads, considering multiple possible `typical' ways to merge the reads, where the typicality will be defined based on the concentration properties of some quantities we will subsequently define.  
    
    For each such typical merge process, we get a set of \textit{islands}, where an island refers to a string of maximal length obtained in the merging process (formal definitions follow in subsequent subsections). 
    Ultimately, upon going through all possible orderings, several such island sets may be generated. In the second phase, which we call the \textbf{filtering phase}, these island sets are then filtered based on further typicality constraints. The filtered island sets which pass the final typicality conditions are referred to as \textit{candidate island sets}. The third and final phase is called the \textbf{compatibility check phase}. In this phase, for each candidate island set, the decoder checks if all the islands of that candidate island set occur as compatible substrings of any codeword. If there is precisely one codeword $\vx_{\hat{w}}$ in $\cal C$ that passes this check, across all the candidate island sets, then the estimate is declared as $\hat{W}=\hat{w}$. Otherwise, a decoding failure is declared. 
   We show that the decoding algorithm results in the correct estimate, i.e., $W=\hat{W}$ with high probability, as $\bl$ grows large. A more precise description and analysis of the decoding is provided in Subsection \ref{subsec:descriptiondecoding} and \ref{subsec:preiciseanalysisofdecoding}. Subsection \ref{subsec:mergingdefns} and Subsection \ref{subsec:Concentrationlemmas} describe the various quantities required for the description and analysis of the decoder, and the concentration results on some of these quantities, respectively. 
\end{itemize}

\subsection{Merging and Coverage:   Definitions and Terminology}
\label{subsec:mergingdefns}
We now give the formal definitions and terminology for various quantities. 
Again, these quantities are either identical or parallel to those defined in \cite{ravi_coded_ssc}. 
\begin{definition}[Length and Size of string]
    For any $\vu\in\{0,1,\eras\}^\star$, the length of $\vu$ is denoted by $\length{\vu}$. The  \textit{size of $\vu$} is the number of unerased bits in $\vu$ and is denoted by $\size{\vu}$.
\end{definition}
\begin{definition}[Prefix and Suffix]
    For a string $\vv\in\{0,1,\eras\}^{l}$ and any positive integer $l'\leq l$, a string $\vz\in\{0,1,\eras\}^{l'}$ is said to be a $l'$-\textit{suffix} of $\vv$, if $(v_{l-l'+1},\hdots,v_{l})=\vz$, and is denoted by $\suff{\vv}{l'}$. Similarly, if $(v_1,\hdots,v_{l'})=\vz$, then $\vz$ is said to be a $l'$-\textit{prefix} of $\vv$ and is denoted by $\pref{\vv}{l'}$.
\end{definition}
\begin{definition}[Compatibility, $l$-compatible strings and substring compatibility]
    Let $\vu$ and $\vv$ be any two strings in $\{0,1,\eras\}^l$. We say that $\vv$ and $\vu$ are \textit{compatible}, if \begin{align*}u_i=\eras~\text{or}~v_i=\eras,~~~
  \text{for any }~ i\in[l]~ \text{s.t.}~u_i\neq v_i.\end{align*}
     For any $\vu, \vv\in \{0,1,\eras\}^\star$ (not necessarily of same length), the string $\vv$ is said to be a \textit{compatible substring} of $\vu$, if $\vv$ is compatible with any substring of $\vu$. Finally, for any $\vu, \vv\in \{0,1,\eras\}^\star$, we say $\vv$ is \emph{$l$-compatible} with $\vu$, if $\suff{\vu}{l}$ and $\pref{\vv}{l}$ are compatible. 
\end{definition}

\begin{figure}[tbh]
\centering
\includegraphics[width=\columnwidth]{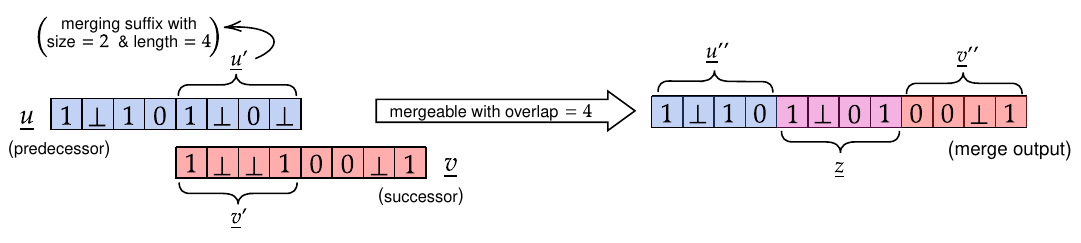}
\caption{
Illustration of a merge operation.}
\label{fig:MergeDiagram}
\end{figure}

We define the merging of two reads in the following manner. 
\begin{definition}[Merge of two strings]
    Let $\vu$ and $\vv$ be any two strings in $\{0,1,\eras\}^\star$ such that $\vv$ is $l$-compatible with $\vu$. Let $\suff{\vu}{l} = \vu'$ and $\pref{\vv}{l} = \vv'$. Suppose $\size{\vu'}\neq 0$. Then, we say that $\vu$ and $\vv$ are \emph{mergeable with overlap} $l$. 
    The output of the merge operation is defined as the string $(\vu''~|~\vz~|~\vv'')$ obtained by the concatenation of three substrings: $\vu''$, $\vz$ and $\vv''$, where $\vu''=\pref{\vu}{\length{\vu}-l}$, $\vv''=\suff{\vv}{\length{\vv}-l}$, and $\vz$ is defined as follows.
    \begin{align*}
        z_i = \begin{cases}
                0 & \text{if } u_{i}' = 0 \text{ or } v_{i}' = 0, \\
                1 & \text{if } u_{i}' = 1 \text{ or } v_{i}' = 1,\\
                \eras & \text{if } u_{i}' = \eras \text{ and } v_{i}' = \eras
            \end{cases}.
    \end{align*}
    With respect to the merge defined above, we term the substring $\vu'$ as the \emph{merging suffix}, or simply the \emph{suffix}. Fig. \ref{fig:MergeDiagram} shows an illustration of this merge operation. 
\end{definition}

We also recall that, during the sequencing process, each of the read $\vy\in\caly$ has a certain starting position $\start{\vy}$. The following terminologies are regarding the ground truth of $\caly$.
\begin{definition}[True Successors, Ordering, and Overlaps]  The \textit{true successor} of a read $\vy_1\in\caly$ is another read $\vy_2\in\caly$, such that $\start{\vy_2}\geq \start{\vy_1}$ (in cyclic wrap around fashion) and $(\start{\vy_2}-\start{\vy_1})$ is smallest among all reads $\vy_2\in \caly\setminus\{\vy_1\}$. Thus, the \textit{true ordering}, is an ordering of the $K$ reads such that each read is succeeded by its true successor. The \textit{true overlap} between any read $\vy_1$ and its true successor $\vy_2$ is defined to be $0$, if $\start{\vy_2}>\start{\vy_1}+L-1$ (in cyclic wrap around manner). If $\start{\vy_1}\leq \start{\vy_2}\leq \start{\vy_1}+L-1$, then the true overlap of $\vy_1$ with $\vy_2$ is $\start{\vy_1}+L-\start{\vy_2}.$ 
\end{definition}

As mentioned before, our algorithm merges the reads corresponding to various orderings and typical overlaps. We now formally define the notion of an island arising out of a merging process of the reads, following a given ordering and a tuple prescribing the sizes of the merging suffixes.

\begin{definition}[Orderings, Islands, and True Islands]
\label{defn:islandsandtrueislands}
    Let $\upzeta$ denote a permutation of $[K]$.  Consider the ordering of the $K$ reads defined by $\upzeta$. With respect to this ordering, the read $\vy_{\upzeta(i)}$ is called the \textit{predecessor} of the read $\vy_{\upzeta(i+1)}$, while $\vy_{\upzeta(i+1)}$ is the \textit{successor} of $\vy_{\upzeta(i)}$.   Consider  $\vomega=(\omega_1,\hdots,\omega_\noofreads)\in[0:\rl]^\noofreads$. For some $j\in[K]$, for some positive integer $l$, suppose that the following conditions hold. 
    \begin{itemize}
        \item $\omega_{j'}>0$ and $\vy_{\upzeta(j')}$ is mergeable with  $\vy_{\upzeta(j'+1)}$ with size of the merging suffix being $\omega_{j'}$, for all $j'\in[j:j+l-1]$ (in cyclic wrap around fashion).
        \item 
        $\omega_{j-1}=\omega_{j+l}=0$.
    \end{itemize}
    Then the string obtained by the merging of the reads $\vy_{\upzeta(j')}:j'\in[j:j+l-1]$ successively with their respective successors, is called an \textit{island}. If $\upzeta^t$ is the true ordering of the reads, and each read $\vy$ is merged with its true successor $\vy'$ (as per $\upzeta^t$) based only on the true overlap, i.e., if $\start{\vy}+L-\start{\vy'}>0$, then the islands so obtained are called \textit{true islands}.
\end{definition} 




Another quantity that will use to check the goodness of our islands is the expected number of unerased bits in them. Towards that end, we have the following two definitions. 
\begin{definition}(A bit being \textit{covered} and \textit{visibly covered}):
The $i^\tth$ bit of $\vx_{W}$, denoted by $x_i$, is said to be \emph{covered} by a read $\vy\in{\caly}$ if $\start{\vy} \in [i-\rl+1:i]$. Further, $x_i$ is said to be \textit{visibly covered} by $\vy$ if it is covered by $\vy$ and further unerased in $\vy$. The bit $x_i$ is said to be covered (visibly covered) by the collection of reads $\caly$ if it is covered (visibly covered, respectively) by at least one read in $\caly$.
%
\end{definition}
%
\begin{definition}[Visible Coverage]
The \textit{visible coverage} denoted by $\Phi_v$ of the collection $\caly$ is defined as the fraction of the bits which are visibly covered, respectively) by the reads in $\caly$.  Thus, 
         $
        \Phi_v \triangleq \frac{1}{\bl} \sum_{i=1}^{\bl} \indicatorRV_{\{ x_{i} \text{ is visibly covered by reads in } {\caly}\}}.
        $
\end{definition}
\begin{remark}
    The notion of visible coverage is essentially identical to that notion of \textit{coverage}, as defined in \cite{ravi_coded_ssc}. To be precise, the coverage (denoted by $\Phi$) for a collection $\caly$ is the fraction of bits which are covered by reads in $\caly$, i.e, $\Phi \triangleq \frac{1}{\bl} \sum_{i=1}^{\bl} \indicatorRV_{\{ x_{i} \text{ is covered by reads in } {\caly}\}}$.
\end{remark}

To analyse the errors the decoder can make while merging, we need to bound the different possible ways a read can be merged with other reads in the set $\caly$. To capture this, we define the quantity $M_{\vz}$.
\begin{definition}[The quantity $ M_{\vz}$]
    For a string $\vz = \{0, 1, \eras\}^l, l \in [\rl]$, the random variable  $M_{\vz}$ is defined as the number of reads in ${\caly}$, which are $l$-compatible with $\vz$ (i.e., which have a $l$-length prefix that is compatible with $\vz$). Thus, 
    \begin{align*}
        M_{\vz} \triangleq \sum_{\vy \in {\caly}} \indicatorRV_{\{\vy \text{ is $l$-compatible with } \vz \}}
    \end{align*}
\end{definition}

Towards assessing the goodness of some overlap tuples, we need the following quantity, denoted by $G(\tau)$.

\begin{definition}[The quantity  $G(\tau)$]
\label{defn:gtaudefinition}
We define\footnote{In this work, the $\tau$ in $G(\tau)$ represents the size of the merging suffix (i.e., the number of unerased bits in the overlapping portion of the predecessor) normalised by $\log{\bl}$, rather than the normalised length of overlap itself, as defined in \cite{ravi_coded_ssc}.} $G(\tau)$ to be the number of reads in $\caly$,  such that for each such read, the size of the merging suffix with its true successor is $\tau \log{\bl}$. Thus, 
\begin{align}
\label{eqn:gtaudefinition}
G(\tau)\triangleq \sum_{i\in[K]}\indicatorRV_{\omega^t_i=\tau\log\bl},
\end{align}
where $\vomega^t=(\omega_1^t,\hdots,\omega_K^t)$ is the sequence of sizes of the true merging suffixes. 
\end{definition}
\subsection{Concentration Results and Bounds on Quantities}
\label{subsec:Concentrationlemmas}
In order to show achievability, we will first prove concentration results for some of the quantities that we have defined. Our first concentration result is on the number of true islands. Note that by definition, erasures do not affect the existence of any true islands. 
Hence, the proof for the following lemma is identical to the proof of Lemma 2 in \cite{ravi_coded_ssc}. 
\begin{lemma}[Concentration of Number of True Islands]
\label{lemma:concentr_num_real_islands}
Let the number of true islands be $\noofreads'$. Thus, for any $\epsilon > 0$
\begin{align*}
    \lim_{\bl\to\infty}\Pr\left(\left|\noofreads'-\noofreads e^{-c} \right| \geq \epsilon \noofreads e^{-c}\right)=0.
\end{align*}
\end{lemma}


We now show the concentration of the visible coverage.
\begin{lemma}[Concentration of Visible Coverage]
\label{lemma:concentrationofcoverage}
        For any $\epsilon>0$, the visible coverage $\Phi_v$ satisfies 
        \begin{align}
             \lim_{\bl\to\infty}\Pr\left(|\Phi_v - (1-e^{-\cover(1-\delta)})| > \epsilon(1-e^{-\cover(1-\delta)})\right) = 0.
        \end{align} 
\end{lemma}
\begin{IEEEproof} We have the following equalities, 
    \begin{align*}
        \expect\left[ \Phi_v \right] & = \frac{1}{\bl} \cdot \expect \left[ \sum_{i=1}^{\bl} \indicatorRV_{\{ x_{i} \text{ is visibly covered by in reads in } {\caly}\}} \right] \\ 
                    & = \Pr(x_{1} \text{ is visibly covered by in reads in } {\caly}) \\ 
                    & = 1 - \Pr(x_{1} \text{ is not visibly covered by in reads in } {\caly}) \\
                    & = 1 - \Pr(x_{1} \text{ is not visibly covered by } \vy_{j}, \forall j \in \{1, \cdots, \noofreads\}) \\
                    & = 1 - \Pr(x_{1} \text{ is not visibly covered by read } \vy_{1})^{\noofreads}\\
                    &= 1 - (1 - \Pr(x_{1} \text{ is visibly covered by read } \vy_{1}))^{\noofreads}.
    \end{align*}
    Now, 
     $   \Pr(x_{1} \text{ is visibly covered by read } \vy_{1}) = \frac{\rl}{\bl} (1-\delta).$
    Hence,
    \begin{align*}
        \expect\left[ \Phi_v \right]    &= 1 - \left( 1- \left( \frac{\rl}{\bl} (1-\delta) \right) \right) ^ {\noofreads} \\
                                    &= 1 - \left( 1- \left( \frac{\rl}{\bl} (1-\delta) \right) \right) ^ { \left( \frac{\cover\bl}{\rl}\right) }\\ 
                                    &= 1 - \left( 1- \left( \frac{\normrl \log \bl}{\bl} (1-\delta) \right) \right) ^ { \left( \frac{cn}{\normrl \log \bl}\right) }.
    \end{align*}
    Let $t = \left(\frac{\bl}{\normrl\log \bl (1-\delta)}\right)$. Hence, we get
    \begin{align}
        \nonumber \lim_{\bl \to \infty} \expect\left[ \Phi_v \right]    &=  \lim_{t \to \infty} \expect\left[ \Phi_v \right]\\ 
    \nonumber &= \lim_{t \to \infty} \left( 1 - \left( 1- \frac{1}{t} \right)^{\left(\cover(1-\delta)t\right)}\right)\\
    &= 1 - \lim_{t \to \infty} \left(\left( 1- \frac{1}{t} \right)^{\left(\cover(1-\delta)t\right)}\right)\\
     &= (1-e^{-\cover(1-\delta)}).
    \end{align}

%
We now use the Chebyshev inequality to complete the rest of the proof. Let $A_i$ denote the event that the $i^{\tth}$ symbol $x_{i}$ in the input sequence  is visibly covered by in reads in ${\caly}$. Let ${\cal P} \triangleq \sum_{i=1}^{\bl} \indicatorRV_{A_i}$ and $\bar{{\cal P}} \triangleq \expect[{\cal P}]$. Note that $\Phi_v = \frac{{\cal P}}{\bl}$ and $\expect[\Phi_v]=\frac{\bar{{\cal P}}}{\bl}$. Thus,
\begin{align}
    \label{eqn:phi_calc01}
    \nonumber \Pr\left(|\Phi_v - (1-e^{-\cover(1-\delta)})| > \epsilon(1-e^{-\cover(1-\delta)})\right) &= \Pr\left(|\Phi_v - \expect[\Phi_v]| > \epsilon \expect[\Phi_v]\right)\\ &= \Pr\left(|{\cal P} - \bar{{\cal P}}| > \epsilon \bar{{\cal P}} \right).
\end{align}

Further, we know that
\begin{align}
    \label{eqn:phi_calc02}
    \Pr\left(|{\cal P} - \bar{{\cal P}}| > \epsilon \bar{{\cal P}} \right) \leq \frac{\Var({\cal P})}{\epsilon^2 \bar{{\cal P}}^2}.
\end{align}

Since ${\cal P}$ is the sum of indicator random variables, the following is true (for a proof, see Chapter 4 in \cite{g_tau_concentration}, for instance),

\begin{align}
\label{eqn:phi_calc03}
    \Var({\cal P}) \leq \bar{{\cal P}} + \sum_{i, j \in [\bl]: i \neq j} \Cov(\indicatorRV_{A_i}, \indicatorRV_{A_j}).
\end{align}
where
\begin{align}
\label{eqn:phi_calc04}
    \Cov(\indicatorRV_{A_i}, \indicatorRV_{A_j}) = \expect[\indicatorRV_{A_i}\indicatorRV_{A_j}] - \expect[\indicatorRV_{A_i}]\expect[\indicatorRV_{A_j}],
\end{align}
is the covariance between the random variables $\indicatorRV_{A_i}$ and $\indicatorRV_{A_j}$.

Now, 
\begin{align}
\label{eqn:phi_calc05}
    \expect[\indicatorRV_{A_i}] = \Pr(A_i) = \frac{\bar{\cal P}}{\bl}.
\end{align}

Further, $A_i$ and $A_j$ are independent if $|i-j|\geq \rl$. Thus,

\begin{align}
\label{eqn:phi_calc06}
    \nonumber \expect[\indicatorRV_{A_i}\indicatorRV_{A_j}] &= \Pr(A_i,A_j)\\
    \nonumber &= \Pr(A_i,A_j | \{|i-j|\geq \rl\}) + \Pr(\{|i-j|<\rl\})\\
    \nonumber &= \Pr(A_i)\Pr(A_j) + \frac{\rl}{\bl}\\
    &= \frac{\bar{\cal P}^2}{\bl^2} + \frac{\rl}{\bl}.
\end{align}
Using  (\ref{eqn:phi_calc05}) and (\ref{eqn:phi_calc06}) in (\ref{eqn:phi_calc03}), we see that $\Cov(\indicatorRV_{A_i}, \indicatorRV_{A_j})=\rl/\bl$. Using this in (\ref{eqn:phi_calc03}), we obtain 
%
\begin{align}
\label{eqn:phi_calc07}
     \Var({\cal P}) \leq \bar{{\cal P}} + \sum_{i, j \in [\bl]: i \neq j} 
     \frac{\rl}{\bl} \leq \bar{{\cal P}} + \bl \rl.
\end{align}

Using (\ref{eqn:phi_calc02}) and (\ref{eqn:phi_calc07}) in (\ref{eqn:phi_calc01}) and the fact that $\bar{{\cal P}}=\bl\expect[\Phi_v ]=\bl(1-e^{-\cover(1-\delta)})$,
\begin{align*}
    \Pr\left(|\Phi_v - (1-e^{-\cover(1-\delta)})| > \epsilon(1-e^{-\cover(1-\delta)})\right) &\leq \frac{\bar{{\cal P}} + \bl \rl}{\epsilon^2 \bar{{\cal P}}^2} = \left(\frac{1}{\epsilon^2 \bar{{\cal P}}}\right) + \left(\frac{\bl \rl}{\epsilon^2 \bar{{\cal P}}^2}\right)\\
     &= \Theta\left(\frac{1}{\bl}\right) + \Theta\left(\frac{\log{\bl}}{\bl}\right) \to 0,
\end{align*}
as $\bl \to \infty$. This completes the proof. 
\end{IEEEproof}
\begin{remark}
\label{remark:concentrationofplaincoverage}
    Note that the concentration of the coverage $\Phi$ (i.e., Lemma 1 in \cite{ravi_coded_ssc}) also follows from Lemma \ref{lemma:concentrationofcoverage}, by substituting $\delta=0$.
\end{remark}

Lemma \ref{lemma:concentrationofgtau} shows the concentration of the parameter $G(\tau)$ in some regime around its mean. Lemma \ref{lemma:concentrationofgtau} is similar to Lemma 4 in \cite{ravi_coded_ssc}, and the proof follows similar arguments. However, a key difference with \cite{ravi_coded_ssc} is that the deviation around the mean in Lemma 4 of \cite{ravi_coded_ssc} is a function of $\gbar(\tau)$, whereas for our purpose here, the deviation ${\epsilon\bl}/{\log^2{\bl}}$ suffices. 
\begin{lemma}
\label{lemma:concentrationofgtau}
 For any $\epsilon > 0$,
    \begin{align*}
         \lim_{\bl\to\infty}\Pr{\left(\bigcup_{\tau \in {\cal T}} \left\{\left|G(\tau)-\gbar(\tau)\right|\geq  \frac{\epsilon\bl}{\log^2{\bl}} \right\} \right)} = 0,
    \end{align*}
    where ${\cal T} = \{ \frac{1}{\log{n}},  \frac{2}{\log{n}}, \cdots, \normrl \}$ and $\gbar(\tau) \triangleq \expect[G(\tau)]$.
\end{lemma}


\begin{IEEEproof}
We know that,
    \begin{align}
    \label{eqn:concentrationofgtau01}
        \Pr{\left(\bigcup_{\tau \in {\cal T}} \left\{\left|G(\tau)-\gbar(\tau)\right|\geq \epsilon \frac{\bl}{\log^2{\bl}} \right\} \right)} \leq \frac{\Var{(G(\tau))}}{\epsilon^2 \cdot \left(\frac{\bl}{\log^2{\bl}}\right)^2}.
    \end{align}

The proof follows similar arguments as in the proof of Lemma \ref{lemma:concentrationofcoverage}, to bound the variance term. Let $A_i$ denote the event that the $i^{\tth}$ read $\vy_i$ overlaps with its true successor with size of merging suffix as $\tau \log{\bl}$. Thus, $G(\tau) = \sum_{i=1}^\noofreads \indicatorRV_{A_i}$. As the random variables $A_i:i\in[K]$ are identically distributed, we have $\gbar(\tau) = \noofreads \Pr(A_i)$, for any $i$.

As $G(\tau)$ is the sum of indicator random variables, the following is true (for a proof, see Chapter 4 in \cite{g_tau_concentration}, for instance),
\begin{align}
\label{eqn:concentrationofgtau02}
    \Var(G(\tau)) \leq \gbar(\tau) + \sum_{i,j \in [\noofreads] : i\neq j} \Cov(\indicatorRV_{A_i}, \indicatorRV_{A_j}),
\end{align}
where
\begin{align}
\label{eqn:concentrationofgtau03}
    \Cov(\indicatorRV_{A_i}, \indicatorRV_{A_j}) = \expect[\indicatorRV_{A_i}\indicatorRV_{A_j}] - \expect[\indicatorRV_{A_i}]\expect[\indicatorRV_{A_j}],
\end{align}
is the covariance between the random variables $\indicatorRV_{A_i}$ and $\indicatorRV_{A_j}$.

Now, 
\begin{align}
\label{eqn:concentrationofgtau04}
    \expect[\indicatorRV_{A_i}] = \Pr(A_i) = \frac{\gbar(\tau)}{\noofreads} = \expect[\indicatorRV_{A_j}] .
\end{align}

Further, $A_i$ and $A_j$ are independent if reads $\vy_i$ and $\vy_j$ do not overlap, i.e., $|\start{\vy_i} - \start{\vy_j}| \geq \rl$. Thus, we have
\begin{align}
\label{eqn:concentrationofgtau05}
    \expect[\indicatorRV_{A_i}\indicatorRV_{A_j}] &= \Pr(A_i, A_j) \nonumber \\
    &\leq\Pr(A_i, A_j| \{|\start{\vy_i} - \start{\vy_j}| \geq \rl\})\nonumber\\ 
    &~~~~+ \Pr(\{|\start{\vy_i} - \start{\vy_j}| < \rl\}\nonumber)\\
    &\leq\Pr(A_i)\Pr(A_j)+\frac{\rl}{\bl} \nonumber\\ 
    &= \frac{\gbar(\tau)^2}{\noofreads^2} +\frac{\rl}{\bl}.
\end{align}

Using (\ref{eqn:concentrationofgtau03}), (\ref{eqn:concentrationofgtau04}) and (\ref{eqn:concentrationofgtau05}) in (\ref{eqn:concentrationofgtau02}), we get
\begin{align}
\nonumber
    \Var(G(\tau)) &\leq \gbar(\tau) +  \sum_{i,j \in [\noofreads] : i\neq j} \frac{\rl}{\bl}\\
    &\leq \gbar(\tau) + \noofreads^2\frac{\rl}{\bl}\\
    \label{eqn:concentrationofgtau08}
    &\stackrel{(a)}{\leq} \noofreads + c\noofreads = (1+c)\noofreads,
\end{align}
where (a) holds as $G(\tau)\leq K$, by definition. Using (\ref{eqn:concentrationofgtau08}) in (\ref{eqn:concentrationofgtau01}), 
\begin{align*}
    \Pr&\left(\bigcup_{\tau \in {\cal T}} \left\{\left|G(\tau)-\gbar(\tau)\right|\geq \epsilon \frac{\bl}{\log^2{\bl}} \right\} \right) \leq \frac{\Var{(G(\tau))}}{\epsilon^2 \cdot \left(\frac{\bl}{\log^2{\bl}}\right)^2}\leq \frac{(1+c)\noofreads}{\epsilon^2 \cdot \left(\frac{\bl}{\log^2{\bl}}\right)^2} =\Theta\left(\frac{\log^3{\bl}}{\bl}\right) \to 0,
\end{align*}
as $\bl \to \infty$.
\end{IEEEproof}
    

    
 Lemma \ref{lemma:Mzconcentration} gives us concentration results for the parameter $M_{\vz}$ in various regimes.  This is similar to Lemma 3 in \cite{ravi_coded_ssc}, except that the proof is simpler than in \cite{ravi_coded_ssc}. 
\begin{lemma}
\label{lemma:Mzconcentration}
    For $\vz\in \cup_{l\in[\rl]}\{0, 1, \eras\}^l$, let $\tau_{ue}(\vz)=\size{\vz}/\log\bl$. The following are then true for any $\epsilon > 0$, 
    \begin{enumerate}[a)] 
        \item 
        $\lim\limits_{\bl\to\infty}\Pr\left( \bigcup_{\vz\colon \tau_{ue}(\vz) \leq 1- \epsilon} \left\{ \left| M_{\vz} - \noofreads n^{-\tau_{ue}(\vz)}\right| \geq \epsilon \noofreads n^{-\tau_{ue}(\vz)} \right\} \right)= 0.
        $
        \item 
        $
            \lim\limits_{\bl\to\infty}\Pr \left( \bigcup_{\vz\colon  \tau_{ue}(\vz) > 1- \epsilon} \{M_{\vz} \geq n^\epsilon\}\right) = 0.
        $
    \end{enumerate}
\end{lemma}

\begin{IEEEproof}
    Consider vectors with erasures of the form $\vz = \{0, 1, \eras\}^l,$ $l \leq \rl$. For the transmitted codeword $\vx\in \{0,1\}^\bl$, let $D \triangleq \{i\in[\bl]\colon (x_i, \cdots, x_{i+l-1}) \text{ is $l$-compatible with } \vz \}$. Therefore, $M_{\vz} = \sum_{i=1}^{\noofreads} \indicatorRV_{\{\start{\vy_i} \in D\}}$. Observe that $\Pr(\indicatorRV_{\{\start{\vy_i} \in D\}}=1)=1/2^{\tau_{ue}(\vz)\log\bl}=n^{-\tau_{ue}(\vz)}$. Further, the random variables $\indicatorRV_{\{\start{\vy_i} \in D\}}:i\in[K]$ are independent, as $\start{\vy_i}:i\in[K]$ are independent random variables. 

    We first prove part a). Consider $\tau_{ue}(\vz) \leq 1 - \epsilon$. 

    \begin{align*}
        \Pr &\left( \bigcup_{\vz: \tau_{ue}(\vz) \leq 1- \epsilon} \left\{ \left| M_{\vz} - \noofreads n^{-\tau_{ue}(\vz)}\right| \geq \epsilon \noofreads \bl ^{-\tau_{ue}(\vz)} \right\} \right)\\
        &\leq \sum_{\vz: \tau_{ue}(\vz) \leq 1- \epsilon} \Pr \left( \left\{ \left| M_{\vz} - \noofreads n^{-\tau_{ue}(\vz)}\right| \geq \epsilon \noofreads \bl ^{-\tau_{ue}(\vz)} \right\} \right)\\
        &\stackrel{(a)}{\leq} \normrl \log{\bl } \cdot \bl ^{\normrl \log{3}} \cdot\\
        &~~~~\max_{\vz: \tau_{ue}(\vz) \leq 1 - \epsilon} \Pr \left( \left\{ \left| M_{\vz} - \noofreads \bl ^{-\tau_{ue}(\vz)}\right| \geq \epsilon \noofreads \bl ^{-\tau_{ue}(\vz)} \right\} \right)\\
        &\stackrel{(b)}{\leq} \normrl \log{\bl } \cdot \bl ^{\normrl \log{3}} \cdot 2e^{-\left(\frac{K\cdot  \bl ^{-\tau_{ue}(\vz)} \epsilon^2}{(1- \bl ^{-\tau_{ue}(\vz)})}\right)}\\
        &\leq  \normrl \log{\bl } \cdot \bl ^{\normrl \log{3}} \cdot 2e^{-\left(\frac{c \cdot \bl ^{1-\tau_{ue}(\vz)} \epsilon^2}{2\rl}\right)}\\
        &\stackrel{(c)}{\leq} \normrl \log{\bl } \cdot \bl ^{\normrl \log{3}} \cdot 2e^{-\left(\frac{c \cdot \bl ^{\epsilon} \epsilon^2}{2\rl}\right)}.
    \end{align*}

    Here $(a)$ holds as $|\{ \vz: \tau_{ue}(\vz) \leq 1- \epsilon \}| \leq \sum_{l=1}^{\rl} 3^l \leq \sum_{l=1}^{\rl} 3^{\rl} \leq \rl \cdot 3^{\rl} = \normrl \log{\bl } \cdot \bl ^{\normrl \log{3}}$. The inequality $(b)$ is from Hoeffding's inequality in Lemma \ref{lem:heoffding1} (which we use with the parameters $X_i= \indicatorRV_{\{\start{\vy_i} \in D\}}$, $N=K$, and $p= n^{-\tau_{ue}(\vz)}$), and $(c)$ follows as $\tau_{ue}(\vz) \leq 1 - \epsilon$. Thus, part a) of the lemma can be seen to be true, from the R.H.S. of the inequality (c). 
    
        
    

    Similarly, for $\tau_{ue}(\vz) > 1 - \epsilon$, we have,

    \begin{align*}
        \Pr\left( \bigcup_{\vz: \tau_{ue}(\vz) > 1- \epsilon} \{ M_{\vz} \geq n^\epsilon \} \right)
        &\leq \normrl \log{\bl } \cdot \bl ^{\normrl \log{3}} \max_{\vz: \tau_{ue}(\vz) > 1 - \epsilon} \Pr \left( \{M_{\vz} \geq \bl ^\epsilon\} \right)\\
         &\stackrel{(a)}{\leq} \normrl \log{\bl } \cdot \bl ^{\normrl \log{3}} \cdot e^{-\left(\frac{\left( \bl ^\epsilon - \noofreads \bl ^{-\tau_{ue}(\vz)} \right)^2}{2\noofreads \bl ^{-\tau_{ue}(\vz)} (1-\bl ^{-\tau_{ue}(\vz)})}\right)}\\
         &\leq \normrl \log{\bl } \cdot \bl ^{\normrl \log{3}} \cdot e^{-\left(\frac{\left( \bl ^\epsilon - (c/\rl) \bl ^{1-\tau_{ue}(\vz)} \right)^2}{2(c/\rl) \bl ^{1-\tau_{ue}(\vz)}}\right)}\\
         &\stackrel{(b)}{\leq} \normrl \log{\bl } \cdot \bl ^{\normrl \log{3}} \cdot e^{-\left(\frac{\left( \bl ^\epsilon - (c/\rl) \bl ^{\epsilon} \right)^2}{2(c/\rl) \bl^{\epsilon}}\right)}
         \\
         &\leq \normrl \log{\bl } \cdot \bl ^{\normrl \log{3}} \cdot e^{-\left(\frac{ \bl ^\epsilon \left(\normrl\log{\bl} - c \right)^2}{2c\normrl\log{\bl} }\right)},
    \end{align*}
where $(a)$ is due to Hoeffding's inequality in Lemma \ref{lem:heoffding2} (with parameters $X_i=  \indicatorRV_{\{\start{\vy_i} \in D\}}$, $ p= n^{-\tau_{ue}(\vz)}$, and $x= n^\epsilon - \noofreads n^{-\tau_{ue}(\vz)}$) and $(b)$ is because  $\tau_{ue}(\vz) > 1 - \epsilon$. From the above R.H.S. expression, it is easy to see that part b) of the lemma holds. 
\end{IEEEproof}


    
\begin{remark}
Observe that the bounds on $M_{\vz}$ obtained from Lemma \ref{lemma:Mzconcentration} are independent on the vector $\vz$ itself, depending instead only on $\size{\vz}=\tau_{ue}(\vz)\log\bl$. 
\end{remark}

\subsection{Decoding Algorithm}
\label{subsec:descriptiondecoding} 
We are now ready to present the decoding algorithm, Algorithm \ref{alg:decoding_algorithm}. Following the outline presented in Subsection \ref{subsec:outline}, we can understand the decoding algorithm in three phases. The first phase attempts mergings of the reads $\caly$, using the suffix-sizes from a special subset of $[0:L]^K$, defined as follows. 
\begin{definition}[Typical Suffix-size tuples $\Omega$]
\label{defn:typicalsuffixsizetuples}
For a tuple $\vomega \in [0,\rl]^\noofreads$ and integer $b\in[0:L]$, let $\mathtt{count}(\vomega, b)$ be the number of times $b$ appears in $\vomega$. For any $\epsilon>0$, we define the set $\Omega$ of \textit{typical suffix-size tuples} as the set of $\vomega \in [0,\rl]^\noofreads$ satisfying the following conditions.
\begin{itemize}
    \item $|\mathtt{count}(\vomega, 0)-\noofreads e^{-c}|\leq \epsilon \noofreads e^{-c}$, and
    \item $|\mathtt{count}(\vomega, \tau)-\gbar(\tau)|\leq\frac{ \epsilon \bl}{\log^2{\bl}} , \forall \tau \in {\cal T}$, where ${\cal T}=\{ \frac{1}{\log{n}},  \frac{2}{\log{n}}, \cdots, \normrl \}$. 
\end{itemize}    
\end{definition}



For each typical suffix-size tuple $\vomega=(\omega_1,\hdots,\omega_K)\in \Omega$, for each permutation $\upzeta$ of $[\noofreads]$ (which we view as an ordering of the $\noofreads$ reads),  Algorithm \ref{alg:decoding_algorithm} attempts to merge the reads such that each value $\omega_i$ is the size of the merging suffix of each read $\vy_{\upzeta(i)}$
The intuition for the first condition in Definition \ref{defn:typicalsuffixsizetuples} follows from the following observation. As mentioned in Definition  \ref{defn:islandsandtrueislands}, in the process of merging, an island is created, with the last read of the island being read $\vy_{\upzeta(i)}$ whenever $\vomega_i=0$. Thus, $\mathtt{count}(\vomega, 0)$ denotes the number of islands generated in the process of merging the reads, if it is successful. This is the merge phase of Algorithm \ref{alg:decoding_algorithm}, which is performed for each ordering $\upzeta$ and each typical suffix-size tuple $\vomega\in\Omega$, corresponding to the steps 4-8.





In the filtering phase, the set $\cal I$ of islands, obtained from a successful merge process, is filtered based on its visible coverage. That is, the total number of unerased bits in the islands obtained, denoted by  $\phi({\cal I})$, is checked (as per step 9) for the following condition (designed based on Lemma \ref{lemma:concentrationofcoverage}).
\begin{align*}
    | \phi({\cal I}) - (1-e^{-c(1-\delta)})| \leq \epsilon (1-e^{-c(1-\delta)}).
\end{align*}
If the above check is passed, then the set $\cal I$ of islands obtained is added to a collection $\mathsf{CI}$ of \textit{candidate island sets} (step 10). 

The third phase is the compatibility check phase, which is done in steps 15-18. Any codewords which are compatible with all the islands of any set $\cal I$ of islands is added into a set $\hat{\cal X}$ of candidate codewords. Finally, in steps 19-21, the estimated message index $\hat{w}\in[2^{nR}]$ is returned, corresponding to the only codeword in the candidate set $\hat{\cal X}$, if that is the case. Else, a failure is declared (steps 22-23).


\begin{algorithm}[hbt!]
\caption{Decoding Algorithm}\label{alg:decoding_algorithm}

\textbf{Input:} Codebook ${\mathcal{C}}= \{\vx_1, \vx_2, \cdots, \vx_{2^{\bl R}}\}$, Reads $\caly$, Typical suffix-size tuples $\Omega$. 

\textbf{Output:} Estimate $\hat{w}$ of the input message, or Failure.

\textbf{Initialize:} Collection of Candidate Islands $\mathsf{CI} \gets \emptyset$.


\For{\textit{each suffix-size tuple } $\vomega \in \Omega$}{
    \For{\textit{permutation } $\upzeta$ \textit{of } $[\noofreads]$}{
        
        \If{$\vy_{\upzeta(i)}$ \textit{and} $\vy_{\upzeta(i+1)}$ \textit{are mergeable with size of merging suffix as} $\omega_i$, $\forall i \in [\noofreads]$}
        {
            Merge reads according to the suffix-size tuple $\vomega$ to form set of islands $\cal I$. 
            
            $\phi({\cal I}) \gets$ number of unerased bits in resulting islands.
            
            \If{$|\phi({\cal I}) - (1-e^{-c(1-\delta)})| \leq \epsilon (1-e^{-c(1-\delta)})$}{
                Add ${\cal I}$ to $\mathsf{CI}$
            }
        }
    }
}

Candidate codewords $\mathcal{\hat{X}} \gets \emptyset$.

\For{\textit{each set of islands} ${\cal I} \in {\mathsf{CI}}$}{
    Insert into $\mathcal{\hat{X}}$, all the $\vx \in \mathcal{C}$ such that all islands in $\cal I$ are compatible substrings of $\vx$.
}

\eIf{$|\mathcal{\hat{X}}| = 1$}{
Estimate $\hat{w} \gets$ message index corresponding to $\vx \in \mathcal{\hat{X}}$. 

\Return $\hat{w}$
}
{\Return $\mathsf{FAIL}$ (decoding failure)}

\end{algorithm}

\subsection{Brief overview of the proof of achievability}
\label{subsec:briefoverviewofproofofachievability}
 We first show that the true ordering $\upzeta^t$ is surely picked by step 5, and the true suffix-size tuple $\vomega^t$ belongs to $\Omega$ (and thus considered in step 4) with high probability for large $n$, following the concentration properties shown in Lemma \ref{lemma:concentr_num_real_islands} and Lemma \ref{lemma:concentrationofgtau}. The set of islands resulting from these will be the set of true islands, which will have visible coverage close to the expected visible coverage, following Lemma \ref{lemma:concentrationofcoverage}. Thus, the true set of islands, with visible coverage close to the expected visible coverage, will pass the check in step 9, and thus will be in the candidate island set with high probability. Thus, the true transmitted codeword $\vx_w$ belongs to the set of candidate codewords $\hat{\cal X}$, with high probability. Finally,  using the concentration lemmas shown in Subsection \ref{subsec:Concentrationlemmas}, we show that $|\hat{\cal X}|=1$ (therefore containing only the true codeword) with high probability, for large $n$, provided $R$ satisfies (\ref{eqn:achievablerates}) in Theorem \ref{thm:main}. The precise arguments of the proof follow in Subsection \ref{subsec:preiciseanalysisofdecoding}. 

\subsection{Detailed Proof of Achievability}
\label{subsec:preiciseanalysisofdecoding}


Our analysis of the decoder's probability of error follows that in \cite{ravi_coded_ssc}. We define the following undesirable events. 
\begin{align*}
    B_1 &= \{\noofreads' > b_1\}, &B_2&= \{\Phi_v < b_2\},
    \\
    B_3 &= \bigcup_{\vz\in{\cal Z}} \{M_{\vz} > b_3(\tau) \}, &B_4&= \bigcup_{\tau \in {\cal T}}\{G(\tau)>b_4(\tau)\},
\end{align*}
where ${\cal Z}=\cup_{l\in[\rl]}\{0,1,\eras\}^l$, and the constants $b_1,b_2,b_3(\tau)$, and $b_4(\tau)$ are defined as follows. 
\begin{align*}
    b_1 &\triangleq  (1+\epsilon) \noofreads e^{-\cover}, ~~~~b_2 \triangleq (1-\epsilon) (1-e^{-c(1-\delta)}),\\
    b_3(\tau) &\triangleq \begin{cases}
                    (1+\epsilon)\bl^{1-\tau} & \text{if } \tau \leq 1 - \epsilon, \\
                    n^\epsilon & \text{if } \tau > 1 - \epsilon, \\
            \end{cases},\\
            b_4(\tau) &\triangleq \gbar(\tau) +  \frac{\epsilon\bl}{\log^2{\bl}}.
\end{align*}
Let $B = \bigcup_{i=1}^{4} B_{i}$.  
Recall that the transmitted message $W$ is chosen uniformly at random. We thus get the following expression for the probability of decoding error. 
\begin{align}
\label{eqn:basicproboferror}
     \Pr(W\neq \hat{W}) = \Pr(W\neq \hat{W}| W = 1) \leq \Pr(W\neq \hat{W}| W = 1, \bar{B}) + \Pr(B),
\end{align}
by the law of total probability. 

Now, for some island set $\cal I\in\mathsf{CI}$, if each island in $\cal I$ is a compatible substring of some codeword $\vx_i$, then we say that \textit{island set $\cal I$ is compatible with $i$}.  The event $\{\hat{W}\neq 1\}$ can occur in one the following ways: (a) no island set in $\mathsf{CI}$ is compatible with $i=1$ (event $E_1$), or (b) some island set in $\mathsf{CI}$ is compatible with $i\in[2:2^{\bl R}]$ (event $E_i$). Thus, we can write, 
\begin{align}
\label{eqn:proboferrorunionbound}
    \Pr(W\neq \hat{W}| W = 1, \bar{B}) + \Pr(B)\leq \Pr(E_1| W = 1, \bar{B})+\sum_{i=2}^{2^{\bl R}}\Pr(E_i| W = 1, \bar{B})+\Pr(B).
\end{align}
Now, from Lemmas \ref{lemma:concentr_num_real_islands}-\ref{lemma:Mzconcentration}, we then have $\lim_{\bl\to\infty}\Pr(B)=0$. As argued in Subsection \ref{subsec:briefoverviewofproofofachievability}, the event $\bar{B_1}\cap \bar{B_2}\cap \bar{B_4}$ ensures the occurrence of the true island set in the $\mathsf{CI}$ set with high probability as $\bl\to \infty$, following Definition \ref{defn:typicalsuffixsizetuples} and steps 4 to 9 of Algorithm \ref{alg:decoding_algorithm}. This true island set is surely compatible with $i=1$, as $W=1$ is the true message in our case. Thus,  $\lim_{\bl\to\infty}\Pr(E_1| W = 1, \bar{B})=0$. 

Let the collection of candidate islands $\mathsf{CI}=\{{\cal I}_1,\hdots,{\cal I}_{|\mathsf{CI}|}\}$. Hence, we have, for $i \in [2:2^{\bl R}]$,
\begin{align}
\nonumber
    &\Pr(E_i| W=1, \bar{B})\\
    \nonumber&= \Pr(\exists {\cal I} \in \mathsf{CI}~\text{s.t.}~{\cal I}~\text{is compatible with}~i~|~W=1, \bar{B})\\
    \label{eqn:proboferrorindividualmessage}
    &\leq \sum_{s=1}^{|\mathsf{CI}|} Pr\left( {\cal I}_s~ \text{is compatible with}~i~|W=1, \bar{B}\right).
\end{align}
Now, recall that the number of islands in ${\cal I}_s$ is at most $b_1$, for any typical suffix-size tuple $\vomega\in \Omega$.  Further, from the condition on $\phi({\cal I}_s)$ in the filtering phase, the visible coverage of ${\cal I}_s$ must be at least $b_2$. Thus, the islands in ${\cal I}_s$ can be arranged in one of at most $n^{b_1}$ orderings, when checking for compatibility with message $i$. Further, for any such ordering, the probability of compatibility is at most ${2^{-\bl b_2}}$, as the bits in the codewords $\vx_1$ and $\vx_i$ are generated independently and uniformly at random (since $i\neq 1$). Thus, the probability of the event that ${\cal I}_s~ \text{is compatible with}~i$ can be bounded as
\begin{align}
\label{eqn:proboferrorindividualmessageindivislandset}
    \Pr\left( {\cal I}_s~ \text{is compatible with}~i~|W=1, \bar{B}\right)\leq \frac{\bl^{b_1}}{2^{\bl b_2}}.
\end{align}
Using (\ref{eqn:proboferrorindividualmessageindivislandset}), (\ref{eqn:proboferrorindividualmessage}), (\ref{eqn:proboferrorunionbound}) and (\ref{eqn:basicproboferror}), we have 
\begin{align*}
   &\Pr(W\neq \hat{W})  \leq 2^{\bl R} \cdot |\mathsf{CI}| \cdot \bl^{b_1} \cdot \frac{1}{2^{\bl b_2}} + o(1)\\
                    &~=2^{\bl R + \log{|\mathsf{CI}|} + b_1 \log{\bl } - \bl b_2} + o(1)\\
                    &~=2^{\bl R + \log{|\mathsf{CI}|} + (1+\epsilon)\noofreads e^{-c} \log{\bl } - \bl (1-\epsilon)\left(1- e^{-c(1-\delta)}\right)}+ o(1).
\end{align*}
Using the fact that $K\log\bl={c\bl}/{\normrl}$, we see that  $\Pr(W\neq \hat{W}) \to 0$ as $n\to \infty$, if
\begin{align}
\nonumber
     R &\leq \lim_{\bl \to \infty} \Big((1-\epsilon)\left(1- e^{-c(1-\delta)}\right) - (1+\epsilon)  \frac{ce^{-c}}{\normrl}-\frac{1}{\bl} \log{|\mathsf{CI}|}\Big)\\
\label{eqn:rateupperboundwithlogCI}
    &\leq (1-\epsilon) \left(1- e^{-c(1-\delta)}\right) - (1+\epsilon)  \frac{ce^{-c}}{\normrl} - \lim_{\bl \to \infty}\frac{1}{\bl} \log{|\mathsf{CI}|}.
\end{align}
In Appendix \ref{app:boundfor1bynlogCI}, the term $\frac{1}{\bl} \log{|\mathsf{CI}|}$ is shown to be upper bounded as shown below, for any $p>0$ and $d>0$. 
\begin{align}
\label{eqn:boundfor1bynlogCI}
    \frac{1}{\bl} \log{|\mathsf{CI}|} &\leq \frac{(1+2\slackforavglambdaue)\normsteplambdaue}{(1-\erasprob)} \cdot\left(\frac{c}{\normrl}\right)^{2} \cdot e^{\alpha \slackforavglambdaue\normsteplambdaue-c}\cdot \left(\frac{e^{\alpha\normsteplambdaue}(e^{\alpha}-1)}{(e^{\alpha\normsteplambdaue}-1)} - \frac{e^{2\alpha \normsteplambdaue}\left(\left(e^{\alpha(1+\normsteplambdaue)} - e^{\alpha}\right) - \normsteplambdaue\left(e^{\alpha(1+\normsteplambdaue)} - 1\right)\right)}{\left(e^{\alpha\normsteplambdaue} - 1\right)^2}\right),
\end{align}
where $\alpha = \frac{c}{\normrl (1-\delta)}$.

Using (\ref{eqn:boundfor1bynlogCI}) in (\ref{eqn:rateupperboundwithlogCI}) and letting $p\to 0$, we get that the condition on $R$ becomes
\begin{align}
\label{eqn:achievableratesintermediate}
 R<\left(1- e^{-c(1-\erasprob)}\right) - \frac{ce^{-c}}{\normrl} - \upbeta(d), 
\end{align}
where 
\begin{align*}
\upbeta(d)= \frac{\normsteplambdaue}{(1-\erasprob)} \cdot\left(\frac{c}{\normrl}\right)^{2} \cdot e^{-c}\cdot \left(\frac{e^{\alpha\normsteplambdaue}(e^{\alpha}-1)}{(e^{\alpha\normsteplambdaue}-1)} - \frac{e^{2\alpha \normsteplambdaue}\left(\left(e^{\alpha(1+\normsteplambdaue)} - e^{\alpha}\right) - \normsteplambdaue\left(e^{\alpha(1+\normsteplambdaue)} - 1\right)\right)}{\left(e^{\alpha\normsteplambdaue} - 1\right)^2}\right).
\end{align*}

Now, as $d \to 0$, 
\begin{align}
    \label{eqn:closedformbetad}
    \lim_{d \to 0} \upbeta(d) = (1-\delta)(e^{-c\left(1-\frac{1}{\normrl (1-\delta)}\right)}-e^{-c}) + \frac{c}{\normrl} e^{-c}.
\end{align}
The proof of (\ref{eqn:closedformbetad}) is provided in Appendix \ref{app:betad}.

Thus, using (\ref{eqn:closedformbetad}) in (\ref{eqn:achievableratesintermediate}), we have proven our result, which is that if 
\begin{align*}
    R<\left(1- e^{-c(1-\erasprob)}\right) - (1-\delta)\left(e^{-c\left(1-\frac{1}{\normrl (1-\delta)}\right)} - e^{-c}\right),
\end{align*}
then $\Pr(W\neq \hat{W}) \to 0$ as $n\to \infty$.

\section{Conclusion} 
\label{sec:conclusion}
In this work, we identified achievable rates for the shotgun sequencing channel $\ssechannel$ with erasure probability $\delta$, using techniques inspired from \cite{ravi_coded_ssc}. 
While a closed-form expression for  the result in Theorem \ref{thm:main} has eluded us so far, the expression for the $\delta=0$ case is identical to the capacity of the erasure-free shotgun sequencing channel, derived in \cite{ravi_coded_ssc}.
As expected, we see that the obtained achievable rate for $\ssechannel$ reduces progressively as $\delta$ increases. For the shotgun sequencing channel, the converse result obtained in \cite{ravi_coded_ssc} depends on results from prior work on the torn-paper channel \cite{shomorony_torn_paper,ravi_torn_paper_lost_pieces}. A converse result for $\ssechannel$ can possibly be obtained by generalizing these results to torn-paper channels with erasures; however, this appears not straightforward, a fact that has been noticed before (see \cite[Section VII]{ravi_torn_paper_lost_pieces}). This is currently a work in progress. Further, our result here is essentially based on the random coding arguments, and are not computationally efficient. Developing efficiently encodable and decodable codes for this channel is a direction for future work.

\appendices 

\section{Concentration inequalities used in this work}
The following Hoeffding-type concentration inequalities are used in this work (see \cite{hoeffding}, for instance). 
\begin{lemma}
    \label{lem:heoffding1}
    For i.i.d. Bernoulli random variables $X_1, X_2 \cdots X_{N}$ with parameter $p$,

    \begin{align*}
        \Pr \left( \left| \frac{1}{N} \sum_{i=1}^{N} X_i - p \right| \geq \epsilon p \right) \leq 2e^{-\frac{Np\epsilon^2}{1-p}}.
    \end{align*}
\end{lemma}

\begin{lemma}
    \label{lem:heoffding2}
    For i.i.d. Bernoulli random variables $X_1, X_2 \cdots X_{N}$ with parameter $p$,

    \begin{align*}
        \Pr \left(  \sum_{i=1}^{N} X_i - Np  \geq x \right) \leq e^{-\frac{x^2}{2Np(1-p)}}.
    \end{align*}
\end{lemma}

\section{Proof of (\ref{eqn:boundfor1bynlogCI}) (bound for $\frac{1}{\bl} \log{|\mathsf{CI}|}$)} 
\label{app:boundfor1bynlogCI}
We start with a simple upper bound on $|\mathsf{CI}|$, following steps 4-11 of Algorithm \ref{alg:decoding_algorithm}. 
\begin{align*}
    |\mathsf{CI}|\leq \sum_{\vomega \in \Omega} (\text{number of read-orderings}~\upzeta~ \text{compatible with}~\vomega),
\end{align*}
where an ordering $\upzeta$ is said to be compatible with a suffix-size tuple $\vomega=(\omega_1,\hdots,\omega_\noofreads
)$ if each read $\vy_{\upzeta(i)}$ is mergeable with its successor $\vy_{\upzeta(i+1)}$ with the specified merging suffix-size $\omega_i$, $\forall i\in[\noofreads]$. 

For any $\vomega\in\Omega$, we now provide the intuition for counting the number of compatible orderings. Consider that an arbitrary read $\vy$ is selected as the first read. Note that, due to the presence of erasures, there may be multiple potential merging suffixes with size $\omega_1$ in this specific read $\vy$. A trivial upper bound for the number of such possible merging suffixes is $\length{\vy}=\normrl \log{\bl}$. Now, suppose we pick a particular merging suffix, $\vz$, such that $\size{\vz} = \omega_1=\tau \log n$, where $\tau=\omega_1/\log n$ .


We know that, for a given $\vz$, $M_{\vz}$ represents the number of reads which are $\length{\vz}$-compatible with $\vz$. In other words, $M_{\vz}$ gives the number reads which are mergeable with the read $\vy$ with $\vz$ as merging suffix. Thus, there are at most $M_{\vz}$ possible successors for $\vy$, such that the compatibility with $\vomega$ is maintained. Note that $M_{\vz} \leq b_3(\tau)$ (due to the assumption that the event $\bar{B}$ occurs). Once the  size of the merging suffix and the successor to the first read are fixed, similar counting arguments hold the second read's merge with its successor. Note that the expected number of times $\tau \log{n}$ appears in $\vomega\in\Omega$ is exactly $G(\tau)$, and $G(\tau) \leq b_4(\tau)$ by the definition of $\Omega$. Also, we observe that, since the ordering and merging process are cyclical, only those orderings where the last read is a valid predecessor of the first read, as per the merge given by the suffix-size tuple $\vomega$, are allowed. Thus, every pick where the successor of the last read is not the first read is not considered in the counting. 

 To summarise, for a fixed $\tau$, the number of possible ways of merging a read, choosing some successor and some suffix with size $\tau \log{\bl}$, is upper bounded by $\normrl \log{\bl} \cdot b_3(\tau)$. Such mergings can occur for $G(\tau)$ reads among the $\noofreads$ reads. Using this, we get

\begin{align*}
    |\mathsf{CI}|    &\leq\sum_{\vomega \in \Omega} \prod_{\tau \in {\cal T}} (\normrl \log{\bl} \cdot b_3(\tau))^{b_4(\tau)}\\
            &\leq(\rl +1)^{\noofreads} \cdot \prod_{\tau \in {\cal T}} (\normrl \log{\bl} \cdot b_3(\tau))^{b_4(\tau)}\\
            &=(\rl +1)^{\noofreads} \cdot \prod_{\tau \in {\cal T}} (\normrl \log{\bl} \cdot b_3(\tau))^{\gbar(\tau) + \epsilon \frac{\bl}{\log^2{\bl}}}\\
           &=(\rl +1)^{\noofreads} \cdot \prod_{\tau \in {\cal T}} (\normrl \log{\bl} \cdot b_3(\tau))^{\gbar(\tau)}\\ 
            &\hspace{3cm}\cdot \prod_{\tau \in {\cal T}} (\normrl \log{\bl} \cdot b_3(\tau))^{\epsilon \frac{\bl}{\log^2{\bl}}}.
\end{align*}
Thus, we have 
\begin{align*}
    \lim_{\bl \to \infty} &\frac{1}{\bl} \log{|\mathsf{CI}|}\\
    &\leq\lim_{\bl \to \infty} \frac{1}{\bl} \log{\left((\rl +1)^{\noofreads} \cdot \prod_{\tau \in {\cal T}} (\normrl \log{\bl} \cdot b_3(\tau))^{\gbar(\tau)}\right)}\\
    &+\lim_{\bl \to \infty} \frac{1}{\bl} \log{\left(\prod_{\tau \in {\cal T}} (\normrl \log{\bl} \cdot b_3(\tau))^{\epsilon \frac{\bl}{\log^2{\bl}}}\right)}.
\end{align*}

Now, 
\begin{align*}
    \lim_{\bl \to \infty}& \frac{1}{\bl} \log{\left(\prod_{\tau \in {\cal T}} (\normrl \log{\bl} \cdot b_3(\tau))^{\epsilon \frac{\bl}{\log^2{\bl}}}\right)}\\
    &=\lim_{\bl \to \infty} \frac{\epsilon}{\log^2{\bl}} \log{\left(\prod_{\tau \in {\cal T}} (\normrl \log{\bl} \cdot b_3(\tau))\right)}\\
    &=\lim_{\bl \to \infty} \frac{\epsilon}{\log^2{\bl}} \log{\left(\prod_{\tau \in {\cal T}} \normrl \log{\bl}\right)}\\
    &~~~~~~+ \lim_{\bl \to \infty} \frac{\epsilon}{\log^2{\bl}} \log{\left(\prod_{\tau \in {\cal T}} b_3(\tau)\right)}\\
    &=\lim_{\bl \to \infty} \frac{\epsilon}{\log^2{\bl}} \sum_{\tau \in {\cal T}}\log{\left( \normrl \log{\bl}\right)}\\
    &~~~~~~+ \lim_{\bl \to \infty} \frac{\epsilon}{\log^2{\bl}} \sum_{\tau \in {\cal T}} \log{ b_3(\tau)}\\
    &\stackrel{(a)}{\leq}\lim_{\bl \to \infty} \frac{\epsilon}{\log^2{\bl}} \sum_{\tau \in {\cal T}}\log{\left( \normrl \log{\bl}\right)}\\
    &~~~~~+ \lim_{\bl \to \infty} \frac{\epsilon}{\log^2{\bl}} \sum_{\tau \in {\cal T}} \log{\bl}\\
    &\stackrel{(b)}{\leq}\lim_{\bl \to \infty} \frac{\epsilon}{\log^2{\bl}} (\normrl \log{\bl} + 1)\log{\left( \normrl \log{\bl}\right)}\\
    &~~~~~+ \lim_{\bl \to \infty} \frac{\epsilon}{\log^2{\bl}} (\normrl \log{\bl} + 1) \log{\bl}\\
    &= 0 + \epsilon \normrl.
\end{align*}
Here, $(a)$ holds as $b_3(\tau) \leq \bl, \forall \tau \in {\cal T}$ and $(b)$ is due to $|{\cal T}| = \normrl \log{\bl} + 1$. Thus, as $\epsilon \to 0$, the value of this term goes to 0.

Hence, we have,
\begin{align*}
    \lim_{\bl \to \infty} &\frac{1}{\bl} \log{|\mathsf{CI}|}\\
    &\leq\lim_{\bl \to \infty} \frac{1}{\bl} \log{\left((\rl +1)^{\noofreads} \cdot \prod_{\tau \in {\cal T}} (\normrl \log{\bl} \cdot b_3(\tau))^{\gbar(\tau)}\right)}\\
   &=\lim_{\bl \to \infty} \frac{\noofreads}{\bl} \log{\left(\rl + 1\right)}
    +\lim_{\bl \to \infty} \frac{1}{\bl} \log{\left(\prod_{\tau \in {\cal T}} (\normrl \log{\bl} )^{\gbar(\tau)}\right)}\\
    &~~~~~+ \lim_{\bl \to \infty} \frac{1}{\bl} \log{\left(\prod_{\tau \in {\cal T}} b_3(\tau)^{\gbar(\tau)}\right)}\\
   &=\lim_{\bl \to \infty} \frac{c}{\normrl\log{\bl}} \log{\left(\normrl \log{\bl} + 1\right)}\\
    &~~~~~+\lim_{\bl \to \infty} \frac{1}{\bl} \log{ (\normrl \log{\bl} )} \sum_{\tau \in {\cal T}} \gbar(\tau)\\
    &~~~~~+ \lim_{\bl \to \infty} \frac{1}{\bl} \log{\left(\prod_{\tau > (1-\epsilon)} \left(\bl^\epsilon\right)^{\gbar(\tau)}\right)}\\
    &~~~~~+ \lim_{\bl \to \infty} \frac{1}{\bl} \log{\left(\prod_{\tau \leq (1-\epsilon)} \left(\bl^{1-\tau}\right)^{\gbar(\tau)}\right)}\\
    &=0 +\lim_{\bl \to \infty} \frac{\noofreads}{\bl} \log{ (\normrl \log{\bl} )} \\
    &~~~~~+ \lim_{\bl \to \infty} \frac{\epsilon}{\bl} \log{\bl}  \sum_{\tau > 1-\epsilon} \gbar(\tau)\\
    &~~~~~+ \lim_{\bl \to \infty} \frac{1}{\bl} \log{\bl} \sum_{\tau \leq 1-\epsilon} (1-\tau) \gbar(\tau)\\
   &\leq\lim_{\bl \to \infty} \frac{c}{\normrl\log{\bl}} \log{ (\normrl \log{\bl} )} \\
    &~~~~~+ \lim_{\bl \to \infty} \frac{\epsilon}{\bl} \log{\bl} \cdot \noofreads\\
    &~~~~~+ \lim_{\bl \to \infty} \frac{1}{\bl} \log{\bl} \sum_{\tau \leq 1-\epsilon} (1-\tau) \gbar(\tau)\\
    &=0 + \frac{\epsilon c}{\normrl}\\
    &~~~~~+ \lim_{\bl \to \infty} \frac{1}{\bl} \log{\bl} \sum_{\tau \leq 1-\epsilon} (1-\tau) \gbar(\tau),
\end{align*}
Thus, as $\epsilon \to 0$, 
\begin{align*}
    \lim_{\bl \to \infty} \frac{1}{\bl} \log{|\mathsf{CI}|} \leq \lim_{\bl \to \infty} \frac{1}{\bl} \log{\bl} \sum_{\tau \leq 1-\epsilon} (1-\tau) \gbar(\tau)
\end{align*}

The following claim gives an upper bound for the quantity $\lim_{\bl \to \infty} \frac{\log{\bl}}{\bl}\sum_{\tau \leq 1-\epsilon} (1-\tau) G(\tau)$. Using this completes the proof. 
\begin{claim}
\label{claim:upperboundgtau}
    Let $\gbar(\tau)$ denotes the expectation of $G(\tau)$, and $\alpha=c/(\normrl(1-\erasprob))$. The following statement is true. 
    \begin{align}
\label{eqn:998}
\lim_{\bl\to\infty}&\frac{\log\bl}{\bl}\sum_{\tau=0}^1(1-\tau)\gbar(\tau) \nonumber 
    \\
    \leq &\frac{(1+2\slackforavglambdaue)\normsteplambdaue}{(1-\erasprob)} \cdot\left(\frac{c}{\normrl}\right)^{2} \cdot e^{\alpha \slackforavglambdaue\normsteplambdaue-c}\cdot \nonumber\\ 
&\left(\frac{e^{\alpha\normsteplambdaue}(e^{\alpha}-1)}{(e^{\alpha\normsteplambdaue}-1)}\right. \nonumber\\
    &- \left. \frac{e^{2\alpha \normsteplambdaue}\left(\left(e^{\alpha(1+\normsteplambdaue)} - e^{\alpha}\right) - \normsteplambdaue\left(e^{\alpha(1+\normsteplambdaue)} - 1\right)\right)}{\left(e^{\alpha\normsteplambdaue} - 1\right)^2}\right), 
\end{align} 
for any $p>0$ and any $d>0$. 
\end{claim} 
\begin{IEEEproof}
Recall that $\vomega^t=(\omega^t_1,\hdots,\omega^t_K)$ denotes the true sizes of the merging suffixes of the reads, taken in the true ordering $\upzeta^t$. 
Thus, $\omega^t_i$ refers to the true suffix-size of the $i^{\tth}$ read $\vy_{\upzeta^t(i)}$. Note that  $\omega^t_i:i\in[K]$ are random variables, depending on the length of the overlapping region, as well as the erasure pattern in the overlapping region. Let $L_i$ denote the length of the overlapping region of read $\vy_{\upzeta^t(i)}$ with read $\vy_{\upzeta^t(i+1)}$. Let $\bar{\omega}^t_i$ denote the expectation of $\omega^t_i$ conditioned on $L_i$, over the randomness of the erasures. Observe that $\bar{\omega}^t_i=L_i(1-\delta)$. Also, we recall that $L_i\leq L$.



We split the interval $[0,1)$ into subintervals of size $\normsteplambdaue$ for some $\normsteplambdaue>0$. Thus, there are $1/\normsteplambdaue$ intervals. For $k\in\{0,1,\hdots,1/d-1\}$, the $\intindex^{\tth}$ interval is denoted as $[\intindex\normsteplambdaue,(\intindex+1)\normsteplambdaue)$. Let $\gbar([\intindex\normsteplambdaue,(\intindex+1)\normsteplambdaue))$ denote the expectation of the number of reads which have suffix-size between $\intindex\normsteplambdaue$ and $(\intindex+1)\normsteplambdaue$, i.e., $\gbar([\intindex\normsteplambdaue,(\intindex+1)\normsteplambdaue))\triangleq \expect[\sum_{i\in[K]}\indicatorRV_{\omega^t_i\in [\intindex\normsteplambdaue\log\bl,(\intindex+1)\normsteplambdaue\log\bl)}]$. Thus, $\gbar([\intindex\normsteplambdaue,(\intindex+1)\normsteplambdaue))\leq \sum_{\tau=\intindex \normsteplambdaue}^{(\intindex+1)\normsteplambdaue}\gbar(\tau)$. Thus, we can write 
\begin{align}
\label{eqn:999}
    \sum_{\tau=0}^1(1-\tau)\gbar(\tau)\leq \sum_{\intindex=0}^{1/\normsteplambdaue-1}(1-\intindex\normsteplambdaue)\gbar([\intindex\normsteplambdaue,(\intindex+1)\normsteplambdaue)). 
\end{align}

We now consider $\gbar([\intindex\normsteplambdaue,(\intindex+1)\normsteplambdaue))$. We can write the following inequalities. Without loss of generality, we assume that the true ordering starts with read $\vy_1$  (i.e., $\upzeta^t(1)=1$), and $\vy_1$ starts at the first position, i.e., $\start{\vy_1} =1$. Let $A_{\intindex}$ denote the event that the $\vy_1$ is mergeable with read $\vy_2$ as its successor, with suffix size $\omega^t_1\in [\intindex\normsteplambdaue,(\intindex+1)\normsteplambdaue)$. Let $D_j$ denote the event that $\start{\vy_j}\geq\start{\vy_2}$. Thus, by the definition of $\gbar([\intindex\normsteplambdaue,(\intindex+1)\normsteplambdaue))$ and because $\omega^t_i: i\in[K]$ are identically distributed, we have 
\begin{align}
\nonumber
   \gbar([\intindex\normsteplambdaue,&(\intindex+1)\normsteplambdaue)) \nonumber\\ 
    \nonumber
    = \noofreads&\Pr\Big(\vy_1~\text{is mergeable with}~\vy_{j'}~\text{with suffix-size}\\\nonumber &~~~~\omega^t_1\in [\intindex\normsteplambdaue,(\intindex+1)\normsteplambdaue), ~\text{for some}~j'\in[\noofreads]\setminus\{1\},\\
    \nonumber
   &~~~~\text{ s.t }~\left\{\start{\vy_{j'}}\leq \start{\vy_j}, \forall j\in[\noofreads]\setminus\{1,j'\}\right\}\Big) \nonumber\\
   \label{eqn:900}
   &\hspace{-0.7cm}= \noofreads(\noofreads-1)\Pr(A_{\intindex}, \{D_j:j\geq 3\}). 
\end{align}



 We recall that $\Pr(\start{\vy_j}=s)=1/\bl$, for any $s\in[\bl]$. Since we assumed that $\start{\vy_1}=1$, thus we see that $\start{\vy_2}=\rl-\overlaplength+1=\rl-\avglambdaue/(1-\erasprob)+1$. Thus, for any $1\leq \valueofavglambdaue\leq \rl$, then $\Pr(\avglambdaue=\valueofavglambdaue)=\Pr(\start{\vy_2}=\rl-\valueofavglambdaue/(1-\erasprob)+1)=1/\bl$. We now intend to bound $\Pr(A_{\intindex}, \{D_j:j\geq 3\})$, as $n\to\infty$, using the fact that if $\avglambdaue$ is concentrated in a small interval, then so is $\lambdaue$.

 Consider some small $\slackforavglambdaue>0$ such that $(1+\slackforavglambdaue\normsteplambdaue)\log{\bl}\leq\rl(1-\erasprob)$ (such a $\slackforavglambdaue$ exists, as $\normrl>1/(1-\erasprob)$). We define the following intervals, $1/\normsteplambdaue$ of them.
\begin{align}
I_k=\begin{cases}
[(\intindex-\slackforavglambdaue)\normsteplambdaue,(\intindex+1+\slackforavglambdaue)\normsteplambdaue] , & \text{if}~1\leq \intindex\leq 1/\normsteplambdaue-1\\
[0,(1+\slackforavglambdaue)\normsteplambdaue],&  \text{if}~\intindex=0.
\end{cases}  
\end{align}

Let $C_k$ denote the event that $\avglambdaue\in I_k$. 
We have that, 
\begin{align}
\label{eqn:1000}
    \Pr(A_{\intindex}, \{D_j:j\geq 3\})=& \Pr(A_{\intindex}, \{D_j:j\geq 3\},C_k)\nonumber \\
    &+\Pr(A_{\intindex}, \{D_j:j\geq 3\},\overline{C_k}). 
\end{align} 
We now show that the term $\Pr(A_{\intindex}, \{D_j:j\geq 3\},\overline{C_k})$ is $\bigo(\log\bl/\bl^2)$. We do this in two parts. 

For $\intindex\geq 1$, we can write 
\begin{align}
\nonumber
&\Pr(A_{\intindex}, \{D_j:j\geq 3\},\{\avglambdaue<(\intindex-\slackforavglambdaue)\normsteplambdaue\log{\bl}\})\\
\nonumber
&=\sum_{\valueofavglambdaue<(\intindex-\slackforavglambdaue)\normsteplambdaue\log{\bl}} \Pr(A_{\intindex}, \{D_j:j\geq 3\},\avglambdaue=\valueofavglambdaue)\\
\label{eqn:1001}
&\leq \sum_{\valueofavglambdaue<(\intindex-\slackforavglambdaue)\normsteplambdaue\log{\bl}} \Pr(A_{\intindex}| \avglambdaue=\valueofavglambdaue)\Pr(\avglambdaue=\valueofavglambdaue).
\end{align}
Now, we can use Hoeffding's inequality\footnote{See \cite{hoeffdings_paper_1963}. The inequality is as follows. For $S$ being the sum of $n$ independent Boolean random variables and any $t>0$, $\Pr(S-\mathbb{E}(S)\geq t)\leq \expon^{-2t^2/n}$} to bound the quantity $\Pr(A_{\intindex}|\avglambdaue=\valueofavglambdaue)$. To see this, observe that when $\avglambdaue=\valueofavglambdaue$, the suffix-size $\lambdaue$ of the merge of $\vy_1$ and $\vy_2$ is the sum of $\overlaplength=\valueofavglambdaue/(1-\erasprob)$ independent Boolean indicator random variables ($1$ indicating erasure, $0$ indicating no-erasure). Therefore, we get, for $\valueofavglambdaue<(\intindex-\slackforavglambdaue)\normsteplambdaue\log{\bl}$
\begin{align*}
    &\Pr(A_{\intindex}| \avglambdaue=\valueofavglambdaue)\\
    &\leq \Pr(\lambdaue\geq \intindex\normsteplambdaue\log{\bl} |\avglambdaue=\valueofavglambdaue)\\
    &=\Pr(\lambdaue\geq \avglambdaue+(\intindex\normsteplambdaue\log{\bl}-\avglambdaue) |\avglambdaue=\valueofavglambdaue)\\
    &\stackrel{(a)}{\leq}2\expon^{-2(\intindex\normsteplambdaue\log{\bl}-\valueofavglambdaue)^2(1-\erasprob)/\valueofavglambdaue}\\
    &\stackrel{(b)}{\leq}  2\expon^{-2(\slackforavglambdaue\normsteplambdaue\log{\bl})^2(1-\erasprob)/((\intindex-\slackforavglambdaue)\normsteplambdaue\log{\bl})}=2\expon^{-2\slackforavglambdaue^2(1-\erasprob)\normsteplambdaue\log{\bl}/(\intindex-\slackforavglambdaue)}\\
    &=\Theta(1/\bl), 
\end{align*}
where (a) holds by the Hoeffding's inequality, and (b) holds as $\valueofavglambdaue<(\intindex-\slackforavglambdaue)\normsteplambdaue\log{\bl}$. Using this in (\ref{eqn:1001}), we get 
\begin{align}
\Pr(&A_{\intindex}, \{D_j:j\geq 3\},\{\avglambdaue<(\intindex-\slackforavglambdaue)\normsteplambdaue\log{\bl}\}) \nonumber\\
 \nonumber
&\stackrel{(a)}{\leq} \frac{(\intindex-\slackforavglambdaue)\normsteplambdaue\log{\bl}}{(1-\erasprob)}\cdot \frac{1}{\bl}\cdot 2\expon^{-\slackforavglambdaue^2(1-\erasprob)\normsteplambdaue\log{\bl}/(\intindex-\slackforavglambdaue)}\\
\label{eqn:1002}
&=\Theta\left(\frac{\log\bl}{\bl^2}\right),
\end{align}
where $(a)$ holds because $\avglambdaue$ takes values in steps of $(1-\erasprob)$, (as $\avglambdaue=\overlaplength(1-\erasprob)$ where $\overlaplength$ takes values in unit steps). 

Using similar arguments, we can show the following for all  $\intindex\in\{0,\hdots,1/\normsteplambdaue-1\}$. 
\begin{align}
\nonumber
\Pr&(A_{\intindex}, \{D_j:j\geq 3\},\{\avglambdaue \geq (\intindex+1+\slackforavglambdaue)\normsteplambdaue\log{\bl}\})\\
\nonumber
&=
\sum_{\valueofavglambdaue\geq (\intindex+1+\slackforavglambdaue)\normsteplambdaue\log{\bl}}\hspace{-0.5cm} \Pr(A_{\intindex}, \{D_j:j\geq 3\},\avglambdaue=\valueofavglambdaue)\\
\nonumber
&=
\sum_{\valueofavglambdaue\geq (\intindex+1+\slackforavglambdaue)\normsteplambdaue\log{\bl}} \hspace{-0.5cm}\Pr(A_{\intindex}, \{D_j:j\geq 3\}|\avglambdaue=\valueofavglambdaue)\Pr(\avglambdaue=\valueofavglambdaue)\\
\nonumber
&\leq \sum_{\valueofavglambdaue\geq (\intindex+1+\slackforavglambdaue)\normsteplambdaue\log{\bl}} \Pr(A_{\intindex}| \avglambdaue=\valueofavglambdaue) \Pr(\avglambdaue=\valueofavglambdaue)\\
\nonumber
&\leq \sum_{\valueofavglambdaue\geq (\intindex+1+\slackforavglambdaue)\normsteplambdaue\log{\bl}} \Big(\Pr(\lambdaue< (\intindex+1)\normsteplambdaue\log{\bl} |\avglambdaue=\valueofavglambdaue)\\
\nonumber &\hspace{5cm}\cdot \Pr(\avglambdaue=\valueofavglambdaue)\Big)\\
\nonumber
&\leq \sum_{\valueofavglambdaue\geq (\intindex+1+\slackforavglambdaue)\normsteplambdaue\log{\bl}} \Big(\Pr(\lambdaue< \avglambdaue-(\avglambdaue-(\intindex+1)\normsteplambdaue\log{\bl})\\
&\hspace{5cm}\mid \avglambdaue=\valueofavglambdaue) \Pr(\avglambdaue=\valueofavglambdaue) \Big) \nonumber\\
\nonumber
&\leq \sum_{\valueofavglambdaue\geq (\intindex+1+\slackforavglambdaue)\normsteplambdaue\log{\bl}} 2\expon^{-2(\valueofavglambdaue-(\intindex+1)\normsteplambdaue\log{\bl})^2(1-\erasprob)/\valueofavglambdaue}\Pr(\avglambdaue=\valueofavglambdaue)\\
\nonumber
&\leq \sum_{\valueofavglambdaue\geq (\intindex+1+\slackforavglambdaue)\normsteplambdaue\log{\bl}}\hspace{-0.5cm}2\expon^{-2(\slackforavglambdaue\normsteplambdaue\log{\bl})^2(1-\erasprob)/((\intindex+1+\slackforavglambdaue)\normsteplambdaue\log{\bl})}\Pr(\avglambdaue=\valueofavglambdaue)\\
\nonumber &\leq 2\expon^{-2\slackforavglambdaue^2\normsteplambdaue\log{\bl}(1-\erasprob)/(\intindex+1+\slackforavglambdaue)} \left(\sum_{\valueofavglambdaue\geq (\intindex+1+\slackforavglambdaue)\normsteplambdaue\log{\bl}} \Pr(\avglambdaue=\valueofavglambdaue)\right) \\
\label{eqn:1003}
&=\Theta\left(\frac{\log\bl}{\bl^2}\right). 
\end{align}

Using (\ref{eqn:1002}) and (\ref{eqn:1003}), we see that 
\begin{align}
    \nonumber 
    &\Pr(A_{\intindex}, \{D_j:j\geq 3\},\overline{C_k})\\
    \nonumber 
   &=\Pr(A_{\intindex}, \{D_j:j\geq 3\},\{\avglambdaue<(\intindex-\slackforavglambdaue)\normsteplambdaue\log{\bl}\})\\
    &~~~+\Pr(A_{\intindex}, \{D_j:j\geq 3\},\{\avglambdaue \geq (\intindex+1+\slackforavglambdaue)\normsteplambdaue\log{\bl}\}) \nonumber\\
    \label{eqn:1005}
   &= \bigo\left(\frac{\log\bl}{\bl^2}\right). 
\end{align}
Now, starting with the first term in the R.H.S. of (\ref{eqn:1000}), we have 
\begin{align}
\nonumber
    &\Pr(A_{\intindex}, \{D_j:j\geq 3\},C_k) \\
    &\leq  \Pr(\{D_j:j\geq 3\},\avglambdaue \in I_k)\\
    \nonumber
    & =\sum_{\valueofavglambdaue\in I_k}\Pr(\{D_j:j\geq 3\}|\avglambdaue=\valueofavglambdaue )\Pr(\avglambdaue=\valueofavglambdaue)\\
    \nonumber
    &=\frac{1}{\bl}\cdot\sum_{\valueofavglambdaue\in I_k}\Pr(\{D_j:j\geq 3\}|\avglambdaue=\valueofavglambdaue )\\
    \nonumber
    &=\frac{1}{\bl}\cdot\sum_{\valueofavglambdaue\in I_k}\Pr(\{D_j:j\geq 3\}|\start{\vy_2}=\rl-\valueofavglambdaue/(1-\erasprob)+1)\\
    \label{eqn:1004}
    &=\frac{1}{\bl}\cdot\sum_{\valueofavglambdaue\in I_k}\left(1-\frac{\rl-\valueofavglambdaue/(1-\erasprob)}{\bl}\right)^{\noofreads-2}. 
\end{align}
Using (\ref{eqn:999}), (\ref{eqn:900}), (\ref{eqn:1000}), (\ref{eqn:1005}) and (\ref{eqn:1004}), we get 
\begin{align}
    \nonumber \frac{\log\bl}{\bl}&\sum_{\tau=0}^1(1-\tau)\gbar(\tau) \\
    \nonumber
   &\leq  \frac{\log\bl}{\bl}\cdot(\noofreads(\noofreads-1))\Big(\sum_{\intindex=0}^{1/\normsteplambdaue-1}(1-\intindex\normsteplambdaue)\cdot\\ 
    \nonumber &\hspace{2.5cm}\frac{1}{\bl}\cdot\sum_{\valueofavglambdaue\in I_k}\left(1-\frac{\rl-\valueofavglambdaue/(1-\erasprob)}{\bl}\right)^{\noofreads-2}\Big)\\ 
    \nonumber & \hspace{1cm}+\frac{\log\bl\cdot K(K-1)}{\bl}\Pr(A_{\intindex},\{D_j:j\geq 3\},\overline{C_k}))\\
    \nonumber
   &\leq \frac{\log\bl}{\bl^2}\cdot(\noofreads(\noofreads-1))\cdot\\ 
    \nonumber &\hspace{0.3cm}\Bigg(\sum_{\intindex=0}^{1/\normsteplambdaue-1}(1-\intindex\normsteplambdaue)\cdot\left(\sum_{\valueofavglambdaue\in I_k}\left(1-\frac{\rl-\valueofavglambdaue/(1-\erasprob)}{\bl}\right)^{\noofreads-2}\right)\Bigg)\\ 
    \nonumber &\hspace{1cm}+\bigo\left(\frac{1}{\bl}\right)\\
    \nonumber
    &\leq \frac{\log\bl}{\bl^2}\cdot(\noofreads(\noofreads-1))\cdot\\ 
    \nonumber &\hspace{0.3cm}\Bigg[\sum_{\intindex=0}^{1/\normsteplambdaue-1}(1-\intindex\normsteplambdaue)\cdot\\ 
    \nonumber &\hspace{0.3cm}\left(\sum_{\valueofavglambdaue\in I_k}\left(1-\frac{(\normrl-(\intindex+1+\slackforavglambdaue)\normsteplambdaue/(1-\erasprob))\log\bl}{\bl}\right)^{\noofreads-2}\right)\Bigg]\\ 
    \nonumber &\hspace{1cm}+\bigo\left(\frac{1}{\bl}\right)\\
    \nonumber
    &\leq \frac{\log\bl}{\bl^2}\cdot \left(\frac{c\bl}{\normrl\log{\bl}}\right)^{2} \Bigg[ \sum_{\intindex=0}^{1/\normsteplambdaue-1}(1-\intindex\normsteplambdaue)\cdot\\
    \nonumber &\cdot\Bigg(\sum_{\valueofavglambdaue\in I_k} \left(1-\frac{(\normrl-(\intindex+1+\slackforavglambdaue)\normsteplambdaue/(1-\erasprob))\log\bl}{\bl}\right)^{\noofreads-2}\Bigg)\Bigg]\\ 
    \nonumber &\hspace{1cm}+\bigo\left(\frac{1}{\bl}\right) \nonumber\\
    &\stackrel{(a)}{\leq} \frac{(1+2\slackforavglambdaue)\normsteplambdaue \log{\bl}}{(1-\erasprob)} \cdot \frac{\log\bl}{\bl^2}\cdot \left(\frac{c\bl}{\normrl\log{\bl}}\right)^{2} \nonumber\\ 
    \nonumber &\hspace{0.3cm}\Bigg[\sum_{\intindex=0}^{1/\normsteplambdaue-1}(1-\intindex\normsteplambdaue)\cdot \\ 
    \nonumber &\hspace{0.5cm}\left(1-\frac{(\normrl-(\intindex+1+\slackforavglambdaue)\normsteplambdaue/(1-\erasprob))\log\bl}{\bl}\right)^{\noofreads-2} \Bigg]\nonumber \\ 
    &\hspace{1cm} +\bigo\left(\frac{1}{\bl}\right)\nonumber\\
    &\nonumber \leq \frac{(1+2\slackforavglambdaue)\normsteplambdaue}{(1-\erasprob)} \cdot\left(\frac{c}{\normrl}\right)^{2}\cdot\\ 
    \nonumber &\hspace{0.3cm}\Bigg[\sum_{\intindex=0}^{1/\normsteplambdaue-1}(1-\intindex\normsteplambdaue)\cdot \\ 
    \nonumber &\hspace{0.5cm}\left(1-\frac{(\normrl-(\intindex+1+\slackforavglambdaue)\normsteplambdaue/(1-\erasprob))\log\bl}{\bl}\right)^{\noofreads-2} \Bigg]\nonumber \\ 
    &\hspace{1cm} +\bigo\left(\frac{1}{\bl}\right)\nonumber\\
    &\nonumber\leq \frac{(1+2\slackforavglambdaue)\normsteplambdaue}{(1-\erasprob)} \cdot\left(\frac{c}{\normrl}\right)^{2}\cdot\\ 
    \nonumber &\hspace{0.3cm}\Bigg[\sum_{\intindex=0}^{1/\normsteplambdaue-1}(1-\intindex\normsteplambdaue)\cdot \\ 
    \nonumber &\hspace{0.5cm}\left(1-\frac{(\normrl-(\intindex+1+\slackforavglambdaue)\normsteplambdaue/(1-\erasprob))\log\bl}{\bl}\right)^{\frac{c\bl}{\bar{L}\log{\bl}}-2}  \Bigg]\nonumber \\ 
    &\hspace{1cm} +\bigo\left(\frac{1}{\bl}\right)\nonumber
\end{align}
where (a) due to the size of interval $I_k$.

Now as $\bl \to \infty$, the above value goes to
\begin{align*}
    &\frac{(1+2\slackforavglambdaue)\normsteplambdaue}{(1-\erasprob)} \cdot\left(\frac{c}{\normrl}\right)^{2} \cdot \sum_{\intindex=0}^{1/\normsteplambdaue-1}(1-\intindex\normsteplambdaue) e^{\frac{-c(\bar{L}(1-\erasprob)-(\intindex+1+\slackforavglambdaue)\normsteplambdaue)}{\bar{L}(1-\erasprob)}} \\ 
    \nonumber &=\frac{(1+2\slackforavglambdaue)\normsteplambdaue}{(1-\erasprob)} \cdot\left(\frac{c}{\normrl}\right)^{2} \cdot e^{-c}\cdot \sum_{\intindex=0}^{1/\normsteplambdaue-1}(1-\intindex\normsteplambdaue) e^{\frac{c((\intindex+1+\slackforavglambdaue)\normsteplambdaue)}{\bar{L}(1-\erasprob)}}
\end{align*}

Taking $\alpha = \frac{c}{\bar{L}(1-\erasprob)}$, we get 
\begin{align*}
    &\frac{(1+2\slackforavglambdaue)\normsteplambdaue}{(1-\erasprob)} \cdot\left(\frac{c}{\normrl}\right)^{2} \cdot e^{\alpha \slackforavglambdaue\normsteplambdaue-c}\cdot \sum_{\intindex=0}^{1/\normsteplambdaue-1}(1-\intindex\normsteplambdaue) e^{\alpha(\intindex+1)\normsteplambdaue}\\
   &= \frac{(1+2\slackforavglambdaue)\normsteplambdaue}{(1-\erasprob)} \cdot\left(\frac{c}{\normrl}\right)^{2} \cdot e^{\alpha \slackforavglambdaue\normsteplambdaue-c}\cdot\\
   &\hspace{2cm}\left( \sum_{\intindex=0}^{1/\normsteplambdaue-1}e^{\alpha(\intindex+1)\normsteplambdaue} - \sum_{\intindex=0}^{1/\normsteplambdaue-1}\intindex\normsteplambdaue e^{\alpha(\intindex+1)\normsteplambdaue} \right)\\
   &= \frac{(1+2\slackforavglambdaue)\normsteplambdaue}{(1-\erasprob)} \cdot\left(\frac{c}{\normrl}\right)^{2} \cdot e^{\alpha \slackforavglambdaue\normsteplambdaue-c}\cdot\\ 
    &\hspace{2cm}\cdot\left( \frac{e^{\alpha\normsteplambdaue}(e^{\alpha}-1)}{(e^{\alpha\normsteplambdaue}-1)} - \normsteplambdaue e^{\alpha \normsteplambdaue}\sum_{\intindex=0}^{1/\normsteplambdaue-1}\intindex e^{\alpha\intindex\normsteplambdaue} \right)\\
   &\stackrel{(a)}{=} \frac{(1+2\slackforavglambdaue)\normsteplambdaue}{(1-\erasprob)} \cdot\left(\frac{c}{\normrl}\right)^{2} \cdot e^{\alpha \slackforavglambdaue\normsteplambdaue-c}\cdot \\ &\hspace{1cm}\left(\frac{e^{\alpha\normsteplambdaue}(e^{\alpha}-1)}{(e^{\alpha\normsteplambdaue}-1)}\right.\\
    &\hspace{1.5cm}- \left. \frac{e^{2\alpha \normsteplambdaue}\left(\left(e^{\alpha(1+\normsteplambdaue)} - e^{\alpha}\right) - \normsteplambdaue\left(e^{\alpha(1+\normsteplambdaue)} - 1\right)\right)}{\left(e^{\alpha\normsteplambdaue} - 1\right)^2}\right),
\end{align*}
where $(a)$ holds because $\sum_{i=0}^{N} iq^{i} = \sum_{i=1}^{N} \sum_{j=i}^{N} q^{j} = N\frac{q^{N+1}}{q-1} - \frac{q^{n+1}-1}{(q-1)^2}$, for any $q\neq 0$.
\end{IEEEproof}

\section{Proof of (\ref{eqn:closedformbetad})(expression for $\lim_{d \to 0} \upbeta(d)$)}
\label{app:betad}

Recall that 
\begin{align*}
\upbeta(d)&= \frac{\normsteplambdaue}{(1-\erasprob)} \cdot\left(\frac{c}{\normrl}\right)^{2} \cdot e^{-c}\cdot \left(\frac{e^{\alpha\normsteplambdaue}(e^{\alpha}-1)}{(e^{\alpha\normsteplambdaue}-1)} - \frac{e^{2\alpha \normsteplambdaue}\left(\left(e^{\alpha(1+\normsteplambdaue)} - e^{\alpha}\right) - \normsteplambdaue\left(e^{\alpha(1+\normsteplambdaue)} - 1\right)\right)}{\left(e^{\alpha\normsteplambdaue} - 1\right)^2}\right)\\
&= \left(\frac{1}{(1-\erasprob)} \cdot\left(\frac{c}{\normrl}\right)^{2} \cdot e^{-c}\right)\cdot \normsteplambdaue\left(\frac{e^{\alpha\normsteplambdaue}(e^{\alpha}-1)}{(e^{\alpha\normsteplambdaue}-1)} - \frac{e^{2\alpha \normsteplambdaue}\left(\left(e^{\alpha(1+\normsteplambdaue)} - e^{\alpha}\right) - \normsteplambdaue\left(e^{\alpha(1+\normsteplambdaue)} - 1\right)\right)}{\left(e^{\alpha\normsteplambdaue} - 1\right)^2}\right)\\
&= \left((1-\delta) \alpha^2 e^{-c}\right)\cdot\left(\frac{d.e^{\alpha d} (e^{\alpha} -1)(e^{\alpha d} -1) + d^2 e^{2\alpha d} (e^{\alpha(1+d)}-1)-d e^{2 \alpha d } (e^{\alpha(1+d)} - e^{\alpha})}{(e^{\alpha d} -1)^2}\right),
\end{align*}
where $\alpha = \frac{c}{\normrl (1-\delta)}$.

We start by observing that $\upbeta(d)$ at $d=0$ gives an indeterminate $\left(\frac{0}{0}\right)$ form, and thus we can apply L'Hôpital's rule twice to obtain the limit \footnote{By observation, it was found that the expression remains in the indeterminate $\left(\frac{0}{0}\right)$ form after applying L'Hôpital's rule once, i.e., after differentiating both numerator and denominator once. Only on applying L'Hôpital's rule twice, i.e., after double-differentiating both numerator and denominator, a determinate expression is obtained.}.

Now, we can re-express $\upbeta(d)$ as,
\begin{align}
    \label{eqn:betad01}
    \upbeta(d) = \upgamma \cdot \frac{\upbeta_1+\upbeta_2+\upbeta_3}{\upbeta_4},
\end{align}
where,
\begin{align*}
    \upgamma &= (1-\delta) \alpha^2 e^{-c}\\
    \upbeta_1 &= d.e^{\alpha d} (e^{\alpha} -1)(e^{\alpha d} -1)\\
    \upbeta_2 &= d^2 (e^{\alpha(1+d)}-1)e^{2\alpha d}\\
    \upbeta_3 &= -d e^{2 \alpha d } (e^{\alpha(1+d)} - e^{\alpha})\\
    \upbeta_4 &= (e^{\alpha d} -1)^2.
\end{align*}

Now, 
\begin{align*}
    \frac{\partial \upbeta_1}{\partial d} &= (e^{\alpha} -1)  (e^{\alpha d}(e^{\alpha d}-1) + \alpha d e^{2\alpha d} + \alpha d e^{\alpha d}(e^{\alpha d}-1))\\
    &= (e^{\alpha} -1) (e^{2\alpha d}(1+2\alpha d) - e^{\alpha d} (1+\alpha d)),
\end{align*}
and,
\begin{align*}
   \frac{\partial^2 \upbeta_1}{\partial d^2} &= (e^{\alpha} -1) (2\alpha e^{2\alpha d}(1+2\alpha d) + 2\alpha e^{2\alpha d} - \alpha e^{\alpha d} - \alpha e^{\alpha d}(1+\alpha d)).
\end{align*}
Hence, 
\begin{align}
    \label{eqn:betad02}
    \lim_{d \to 0} \frac{\partial^2 \upbeta_1}{\partial d^2} = (e^{\alpha} -1) (2 \alpha + 2 \alpha - \alpha - \alpha) = 2 \alpha (e^{\alpha} -1). 
\end{align}

Further,
\begin{align*}
    \frac{\partial \upbeta_2}{\partial d} &= 2d (e^{\alpha (1+d)}-1) e^{2\alpha d} + \alpha d^2 e^{\alpha (1+d)} e^{2 \alpha d} + 2 \alpha d^2 (e^{\alpha (1+d)} - 1) e^{2 \alpha d}\\
    &= d e^{2\alpha d} (e^{\alpha (1+d)} (2+3\alpha d) - 2 (1+\alpha d)),
\end{align*}
and,
\begin{align*}
    \frac{\partial^2 \upbeta_2}{\partial d^2} &= e^{2\alpha d} (e^{\alpha (1+d)} (2+3\alpha d) - 2 (1+\alpha d)) + d \cdot  \frac{\partial}{\partial d} (e^{2\alpha d} (e^{\alpha (1+d)} (2+3\alpha d) - 2 (1+\alpha d))).
\end{align*}
Hence,
\begin{align}
    \label{eqn:betad03}
    \lim_{d \to 0} \frac{\partial^2 \upbeta_2}{\partial d^2} = 2(e^\alpha -1).
\end{align}

Furthermore,
\begin{align*}
    \frac{\partial \upbeta_3}{\partial d} &= - (e^{2 \alpha d } (e^{\alpha(1+d)} - e^{\alpha}) + 2\alpha d e^{2 \alpha d } (e^{\alpha(1+d)} - e^{\alpha}) + \alpha d e^{\alpha(1+3d)})\\
    &= - (e^{2\alpha d} (e^{\alpha(1+d)} - e^{\alpha}) (1+2\alpha d) + \alpha d e^{\alpha(1+3d)}),
\end{align*}
and,
\begin{align*}
    \frac{\partial^2 \upbeta_3}{\partial d^2} =  - (&2\alpha e^{2\alpha d} (e^{\alpha(1+d)} - e^{\alpha}) (1+2\alpha d) + \alpha e^{\alpha(1+3d)} (1+2d)\\
    &+ 2\alpha e^{2\alpha d} (e^{\alpha(1+d)} - e^{\alpha}) + \alpha e^{\alpha(1+3d)} + 3 \alpha^2 d e^{\alpha(1+3d)}).
\end{align*}
Hence,
\begin{align}
    \label{eqn:betad04}
    \lim_{d \to 0} \frac{\partial^2 \upbeta_3}{\partial d^2} = -(\alpha e^\alpha + \alpha e^\alpha) = -2\alpha e^\alpha.
\end{align}

Also,
\begin{align*}
    \frac{\partial \upbeta_4}{\partial d} = 2\alpha(e^{\alpha d} - 1)e^{\alpha d},
\end{align*}
and,
\begin{align*}
    \frac{\partial^2 \upbeta_4}{\partial d^2} = 2\alpha^2e^{2\alpha d} + 2\alpha^2(e^{\alpha d} -1)e^{\alpha d}.
\end{align*}
Hence, 
\begin{align}
    \label{eqn:betad05}
    \lim_{d \to 0} \frac{\partial^2 \upbeta_4}{\partial d^2} = 2\alpha^2.
\end{align}

Thus, using (\ref{eqn:betad02}), (\ref{eqn:betad03}), (\ref{eqn:betad04}), and (\ref{eqn:betad05}) in (\ref{eqn:betad01}), 
\begin{align*}
    \lim_{d \to 0} \upbeta(d) &= (1-\delta) \alpha^2 e^{-c}\cdot \left(\frac{2 \alpha (e^{\alpha} -1)+2(e^\alpha -1)-2\alpha e^\alpha}{2 \alpha^2}\right)\\
    &= (1-\delta) e^{-c} (e^\alpha - (1+\alpha))\\
    &=(1-\delta) e^{-c} \left(e^{\frac{c}{\normrl (1-\delta)}} - \left(1+\frac{c}{\normrl (1-\delta)}\right)\right)\\
    &=(1-\delta)(e^{-c\left(1-\frac{1}{\normrl (1-\delta)}\right)}-e^{-c}) + \frac{c}{\normrl} e^{-c}.
\end{align*}

\section*{Acknowledgment}

Hrishi Narayanan thanks IHub-Data, IIIT Hyderabad for extending research fellowship.


\bibliographystyle{IEEEtran}
\bibliography{ISIT24_ShotgunErasures_ArXiv.bib}


\end{document}